\documentclass[twocolumn,prl,superscriptaddress,showpacs,letter]{revtex4-2}
\usepackage{graphicx,nicefrac}
\usepackage{amssymb}
\usepackage{amsfonts}
\usepackage{xcolor}
\usepackage{amsmath}
\usepackage{siunitx}

\newcommand{\rvec}{\vec{r}}
\newcommand{\abs}[1]{\left\lvert #1 \right\rvert}
\renewcommand{\vec}{\mathbf}

\begin{document}

\title{Spectrum of collective excitations of a quantum fluid of polaritons}

\author{F. Claude}
\affiliation{Laboratoire Kastler Brossel, Sorbonne Universit\'{e}, ENS-Universit\'{e} PSL, Coll\`{e}ge de France, CNRS, 4 place Jussieu, 75252 Paris Cedex 05, France}
\author{M. J. Jacquet}
\affiliation{Laboratoire Kastler Brossel, Sorbonne Universit\'{e}, ENS-Universit\'{e} PSL, Coll\`{e}ge de France, CNRS, 4 place Jussieu, 75252 Paris Cedex 05, France}
\author{I. Carusotto}\affiliation{INO-CNR BEC Center and Dipartimento di Fisica, Universit\`{a} di Trento, via Sommarive 14, I-38123 Trento, Italy}
\author{Q. Glorieux}
\affiliation{Laboratoire Kastler Brossel, Sorbonne Universit\'{e}, ENS-Universit\'{e} PSL, Coll\`{e}ge de France, CNRS, 4 place Jussieu, 75252 Paris Cedex 05, France}
\author{E. Giacobino}
\affiliation{Laboratoire Kastler Brossel, Sorbonne Universit\'{e}, ENS-Universit\'{e} PSL, Coll\`{e}ge de France, CNRS, 4 place Jussieu, 75252 Paris Cedex 05, France}
\author{A. Bramati}
\affiliation{Laboratoire Kastler Brossel, Sorbonne Universit\'{e}, ENS-Universit\'{e} PSL, Coll\`{e}ge de France, CNRS, 4 place Jussieu, 75252 Paris Cedex 05, France}

\begin{abstract}
We use a recently developed high-resolution coherent probe spectroscopy method to investigate the dispersion of collective excitations of a polaritonic quantum fluid.
We measure the dispersion relation with high energy and wavenumber resolution, which allows to determine the speed of sound in the fluid, and to evidence the contribution of an excitonic reservoir.
We report on the generation of collective excitations at negative energies, on the ghost branch of the dispersion curve. Precursors of dynamical instabilities are also identified.
Our methods open the way to the precise study of quantum hydrodynamics of quantum fluids of light.
\end{abstract}

\maketitle
.

\section{Introduction}

Quantum fluids of light, made of photons propagating in a nonlinear medium have attracted a lot of interest in the past decade, since their evolution is quite similar to that of atomic Bose-Einstein condensates (BEC) or superfluid liquids ~\cite{london_-phenomenon_1938,Landau1941}. Both BEC and fluid of light systems can be described by analogous equations -- a generalized Gross-Pitaevski equation. Their quantum hydrodynamic properties come from the properties of the excitation modes in the fluid, which are described by the Bogoliubov theory ~\cite{Bogolyubov:1947zz}.
Interestingly, fluids of light are not in equilibrium, and corresponding terms have to be added in the Gross-Pitaevski equation to account for this property. As a result they show both standard quantum fluid properties, like superfluidity and sonic dispersion and a rich ensemble of new behaviors revealed by their collective excitation spectra that we study here.

The quantum fluid of light investigated in this article is a polariton fluid that results from the strong coupling of photons and excitons in a semiconductor microcavity.

Under the condition for cavity resonance, the component of the photon wavevector perpendicular to the cavity $k_{\bot}$ is quantized.
The frequency dependence of photonic modes with wavevector $k$ in the cavity plane is then $\omega=c/n \sqrt{k_{\bot}^2+k^{2}}\approx c/n~ (k_{\bot} + k^{2}/2k_{\bot})$, where $n$ is the refractive index of the cavity.
Thus, the photon frequency acquires a quadratic dependence on $k$ for the motion in the cavity plane (see Fig.~\ref{fig:fig1} (a)) -- cavity photons are endowed with an effective mass $m=\hbar k_{\bot}n/c$ .
Meanwhile, excitons are bound electron-hole states created by laser illumination, i.e. massive interacting excitations.

The strong coupling of photons and excitons generates two new eigenstates of the system: two polariton branches with different dispersive properties, the upper and the lower polaritons, which are shown in Fig.~\ref{fig:fig1} (a).

Polaritons inherit the properties of cavity photons and excitons and thus behave collectively (they interact via the exciton-exciton coupling) as a flow of massive particles (their mass is a combination of the effective photon mass with the exciton mass), that is as a quantum fluid~\cite{carusotto_quantum_2013}.
Similarly to ensembles of cold atoms~\cite{jin_collective_1996,mewes_collective_1996,onofrio_observation_2000,steinhauer_excitation_2002}, polariton fluids exhibit Bose-Einstein condensation ~\cite{kasprzak_boseeinstein_2006,balili_bose-einstein_2007} and superfluidity~\cite{lagoudakis_quantized_2008,utsunomiya_observation_2008,amo_superfluidity_2009,kohnle_single_2011}.
The latter phenomenology is connected to the dispersion of collective excitations of quantum fluids (\textit{eg} Bogoliubov excitations on top of a coherent polariton fluid)
that goes from a parabolic dependence of the energy on the wavenumber $k$ at large $k$ to a linear dependence at low $k$.
While superfluidity phenomenology has been observed over ten years ago in semiconductor microcavities illuminated by resonant laser light, the precise measurement of the spectrum of collective excitations has only been addressed systematically more recently ~\cite{stepanov_dispersion_2019,pieczarka_observation_2020,claude_2022}.

Due to the cavity decay rate and to the exciton finite lifetime, the polaritons generate photoluminescence, which has been used in Refs.~\cite{stepanov_dispersion_2019,pieczarka_observation_2020} to measure the spectrum of Bogolioubov excitations in polariton fluids.
In \cite{pieczarka_observation_2020} the pump laser is far off-resonance and generates a high-energy incoherent exciton reservoir.
In turn, its relaxation toward lower energies leads to the spontaneous generation of a polariton BEC.
In this case, additional exciton-polariton interaction channels are excited~\cite{amelio_perspectives_2020} and, as a result, the shape of the spectra may differ nontrivially from that of the pure polariton condensate in the low-energy and low-wavenumber region, yielding additional effects.
In order to concentrate on the pure polariton quantum fluid properties, the excitation of these additional channels can be decreased by pumping the polaritons on- or near-resonance, thus generating a polariton fluid whose hydrodynamics are directly set by the properties of the pump, as in~\cite{stepanov_dispersion_2019,claude_2022}.

In these configurations, the presence of the high-density pump at low wave number hinders the observation of the luminescence in the low-energy region of the Bogoliubov dispersion.
This can be mended by filtering out the pump component by means of heterodyne four-wave mixing~\cite{kohnle_four-wave_2012} or spectral and polarisation filtering techniques~\cite{stepanov_dispersion_2019}. This allows to access the sonic behavior, obtain the speed of sound in the fluid and highlight the indirect generation of a dark excitonic reservoir.
Yet, the spectral resolution of the experiments~\cite{stepanov_dispersion_2019,pieczarka_observation_2020,kohnle_four-wave_2012} is ultimately limited by the bandwidth of the spectrometer, which is typically larger than the polariton linewidth ~\cite{sermage_lifetime_1996,Bloch_lifetime_1997}.
Therefore, the shape of the dispersion, which strongly depends on the energy and intensity of the pump~\cite{ciuti_quantum_2005}, may not be well resolved, potentially hiding narrow features.

Higher resolution can be achieved by probing the Bogoliubov excitations with an additional laser field with adjustable energy and wavenumber. This versatile spectroscopic method, developed in Ref.~\cite{claude_2022} allows a very precise determination of the dispersion curve in various conditions.
In this paper, we make use of this technique to perform a detailed study of the spectrum of collective excitations on both the normal and ghost branches, and we study the evolution of the Bogolioubov spectrum along the bistability loop of the fluid, as the pump parameters are scanned. We also study the effects of a dark excitonic reservoir~\cite{stepanov_dispersion_2019} on the speed of sound in the fluid and we compare it with the theoretical predictions. 
Finally, at low pump power, we observe plateaus at the crossing of the normal and ghost branches, highlighting the onset of dynamical instability precursors in the fluid. 

\section{Theoretical background}

\subsection{Bogoliubov method}

The evolution equation of the macroscopic wavefunction $\psi(\rvec, t)$ of a dilute Bose-Einstein condensate of ultracold atoms is given by the Gross-Pitaevski equation (GPE) \cite{carusotto_quantum_2013}: 
\begin{equation}
 i\hbar\frac{\partial\psi}{\partial t}= -\frac{\hbar^2}{2m}\nabla^2 \psi+ \hbar g|\psi|^2 \psi+ V_{ext}(\rvec)  \psi,
 \label{GPE}
\end{equation}
with $g$ the interaction constant between the particles and $V_{ext}(\rvec)$ the external trapping potential. The macroscopic fraction of particles occupying the condensate mode behaves collectively: its behaviour can be accurately described within the mean-field approximation, whereby the quantum operator of the atomic matter field $\hat{\psi}(\rvec, t)$ is replaced by the classical wave $\psi(\rvec, t) = \psi_0(\rvec)e^{-i \mu t / \hbar}$. Its phase oscillates at the frequency set by the chemical potential $\mu$, equal to the energy of the condensate, related in free space to the spatially uniform density $\abs{\psi_0(\rvec)}^2$ via the equation of state $\mu = \hbar g n$.

The evolution of small perturbations $\delta\psi(r,t)$ on top of the condensate can be calculated with the Bogoliubov theory ~\cite{Bogolyubov:1947zz}, which consists in looking for the linearized response of the system by setting $\psi_(\rvec,t) = (\psi_0(\rvec) + \delta\psi(\rvec,t))e^{-i\mu t/\hbar}$ in  Eq.(\ref{GPE}). This leads to the Bogoliubov dispersion relation

\begin{eqnarray}
 \omega_B(k) &=& \pm \sqrt{\left(\frac{\hbar k^2}{2m} + gn \right)^2 - \left(gn \right)^2}\\ \label{Bog_disp_v1} &=& \pm\sqrt{\frac{\hbar k^2}{2m}\left(\frac{\hbar k^2}{2m}+ 2gn\right)},
  \label{Bog_disp_v2}
\end{eqnarray}
associating the energy $\hbar \omega_B(k)$ of the weak perturbation (measured from the pump frequency $\omega_0$) to its wavenumber $k$. The + (-) solution is called the positive (negative) Bogoliubov branch, or normal (ghost) branch.

Depending on the value of $k$ compared to  the inverse of the healing length $\xi = \sqrt{\hbar/mgn}$, the collective excitations of the fluid behave in fundamentally different ways. When $k \xi \gg 1$, the dispersion relation has the standard single-particle parabolic shape, 
\begin{equation}
    \nonumber
    \omega_B(k) = \pm\left(\frac{\hbar k^2}{2m}+ gn\right),
\end{equation}
while when $k \xi \ll 1$, it is linear, giving rise to a sonic-like behaviour,
\begin{equation}
    \nonumber
    \omega_B(k) = \pm\ c_s k,
\end{equation}
where the sound velocity is defined as $c_s=\sqrt{\hbar gn/m}$.

\subsection{Out-of-equilibrium case}

The Bogoliubov approach can be extended to polariton fluids, provided that one accounts for the driven-dissipative dynamics in semiconductor microcavities. Indeed, in contrast to atomic condensates, polaritons are intrinsically out-of-equilibrium systems, whose decay rate $\gamma$ has to be compensated by continuous and quasi-resonant pumping $\mathcal{F} (\rvec, t) = F_0(\rvec) e^{-i \omega_0 t}$. The time-dependent evolution of the mean-field wave function is provided by a generalized form of the GPE written as \cite{carusotto_quantum_2013}
\begin{eqnarray}
    i\hbar\frac{\partial\psi_{LP}}{\partial t} &=& \left(\hbar\omega_{LP} -\frac{\hbar^2}{2m_{LP}}\nabla^2+ \hbar g|\psi_{LP}|^2 -i\hbar\frac{\gamma}{2}\right)\psi_{LP} \nonumber\\
    &+& i\hbar\eta(k) F_0,
    \label{GPE1}
 \end{eqnarray}
where $\omega_{LP}$ is the frequency of polaritons, $\left(\hbar^2/2m_{LP}\right)\nabla^2$ gives their kinetic energy (referred as their flow) along the cavity plane, $m_{LP}$ their effective mass, and $\hbar g$ their interaction strength, which is positive (repulsive interaction) in the case where the excitons are generated in a single circular spin state $\sigma$. The coefficient $\eta (k)$, proportional to the transmission of the optical cavity input mirror, quantifies the coupling efficiency of the driving field to polaritons. 

\begin{figure}[!ht]
        \includegraphics[width=1\linewidth]{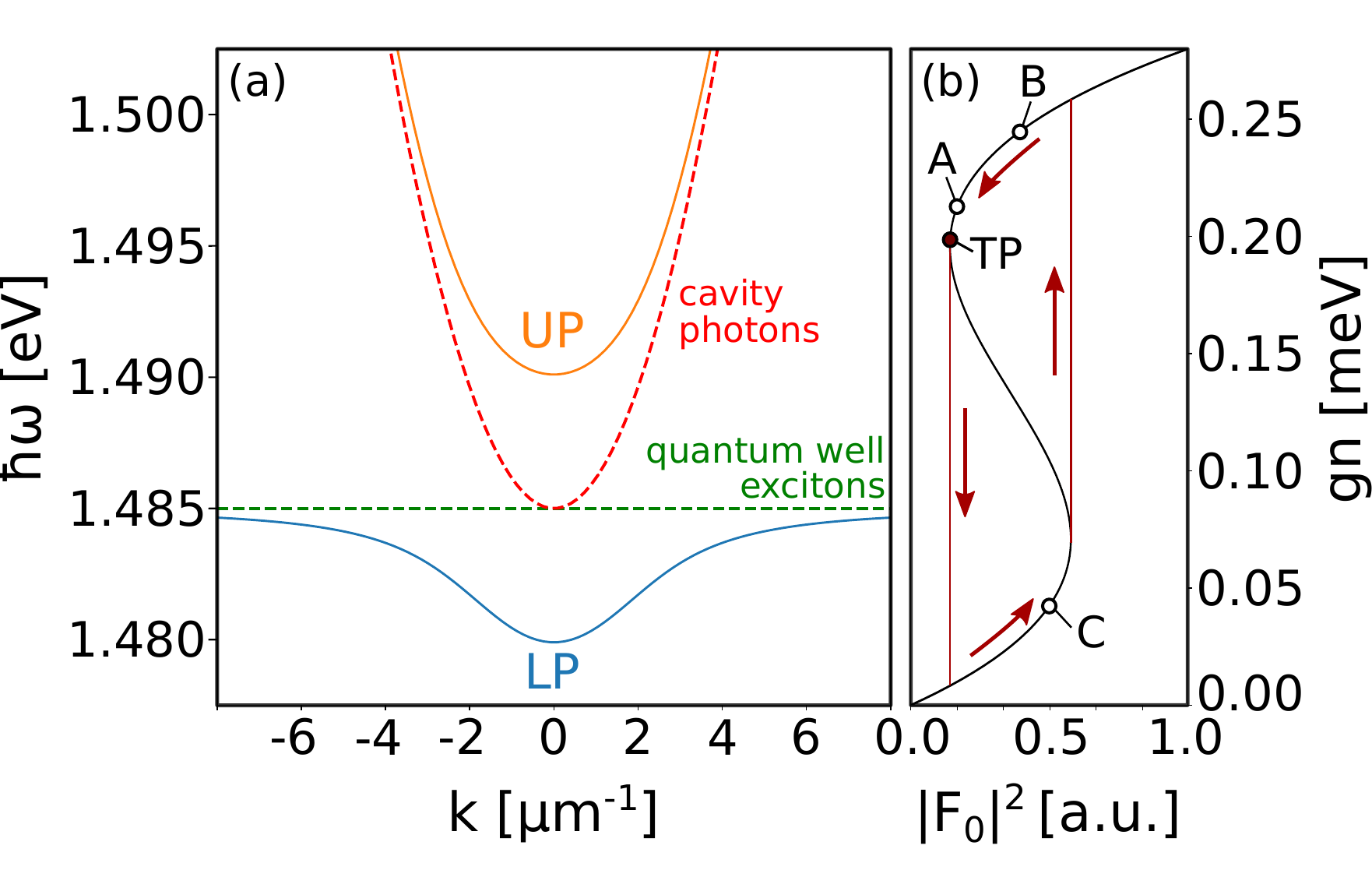}
        \caption{\textbf{Properties of microcavity polaritons} \textbf{(a)} Energy dispersion of the upper (UP) and lower (LP) polaritons. \textbf{(b)} Bistability curve at a fixed pump frequency, indicating the working points A, B, C. The red dot highlights the turning point (TP) of bistability }
    \label{fig:fig1}
\end{figure}

In our experiments, the laser excites the cavity in the vicinity of the lower polariton branch $\omega_0 \simeq \omega_{LP}$ (Fig.~\ref{fig:fig1}). Since the Rabi energy $\hbar \Omega_R$ is much larger than the characteristic energies of the system, in particular the interaction energy $\hbar g n$, it can be assumed that the fluid contains only lower polaritons, allowing to truncate its wave function to that of the lower branch, $\psi(\rvec, t) = \psi_{LP} (\rvec, t)$, without loss of generality.

Moreover, the pump term explicitly breaks the rotational invariance of the global phase of the condensate that prevailed in the equilibrium case. Accordingly, the polariton fluid is not a manifestation of a spontaneous symmetry breaking and therefore it is a very peculiar type  of condensate.  Crucially, rather than being governed by the chemical potential $\mu$, the oscillation frequency of the polariton wave function is fixed by the pump frequency $\psi_{LP}(\rvec, t) = \psi_{LP}^0(\rvec)e^{-i \omega_0 t}$. Therefore, a polariton analogue of the equation of state can be derived in terms of the steady-state generalised GPE. In the case of a pump laser at normal incidence, i.e. with zero wave number, the equation is 
\begin{equation}
n(\rvec) \left[ \left(\frac{\gamma}{2} \right)^{2}+\left( gn - \delta \right) ^{2}\right] = \eta^{2}|F_0(\rvec)|^{2},
\label{bistab}
\end{equation}
which relates the polariton density $n=\abs{\psi_{LP}}^2$ to the laser intensity $I = \abs{F_0(\rvec)}^{2}$ and the frequency detuning $\delta = \omega_0 - \omega_{LP}$, expressed with respect to the bare lower polariton frequency. Remarkably, when $\delta > \sqrt{3}\gamma/2$, the system exhibits a bistable behaviour \cite{baas_bista_2004}, similarly to Kerr non-linear media in an optical cavities. The plot of $n = f\left(I \right)$ in Fig.~\ref{fig:fig1}(b) features a hysteresis loop, where the fluid can either be in a high or low density regime, corresponding to a high or low interaction regime respectively.
The intermediate negative slope density branch is an unstable solution of the generalized GPE that cannot be observed in the standard experimental steady state regime (however, see Ref~\cite{hakim_metamorphosis_2022} for phenomenology beyond the standard case).

This dependence of the polariton density on the pump parameters induces a radical transformation of the collective excitations of polaritons. The evolution of small perturbations on top of a stationary state fluid can be studied by linearizing the generalized GPE with $\psi_{LP}(\rvec, t) = \left[\psi_{LP}^0(\rvec) + \delta \psi_{LP}(\rvec, t) \right] $. In a spatially homogeneous geometry, with a pump at normal incidence with respect to the cavity plane, this leads to the Bogoliubov dispersion relation \cite{carusotto_quantum_2013,ciuti_quantum_2005}:
\begin{eqnarray}
  \omega_B(k) &=& \pm\sqrt{\left(\frac{\hbar k^2}{2m_{LP}} + 2 gn - \delta \right)^2 - (gn)^2} \nonumber\\
  &-& i\gamma/2.
  \label{eq:Bog_disp1}
\end{eqnarray}
where $k$ is the wave number of the perturbation. This equation is similar to Eq.~(\ref{Bog_disp_v1}) for the equilibrium case, except for the additions of an extra frequency shift term $gn - \delta$, coming from the quasi-resonant driving, and of an imaginary term $i \gamma /2$, coming from the polariton finite lifetime. It again exhibits two branches with opposite sign, here called the normal and ghost branches.

When the fluid is injected at the bistability turning point (TP), $\delta = gn$ and thus the real part of the Bogoliubov dispersion relation is exactly the same as in Eq.~(\ref{Bog_disp_v1}). Consequently, we recover the phenomenology of the equilibrium condensates: when $k\xi \ll 1$, the collective excitations behave like phonons, with a speed of sound given by  
\begin{equation}
 c_s=\sqrt{\dfrac{\hbar gn}{m_{LP}}
 } = \sqrt{\dfrac{\hbar \delta}{m_{LP}}}
 \label{eq:sound}
 \end{equation}
 and the polariton fluid may exhibit superfluidity according to the Landau criterion, as was experimentally observed in Refs~\cite{amo_superfluidity_2009,kohnle_single_2011}.
 On the other hand, when $k\xi \gg 1$ the collective excitations behave like single-particles.
 
\subsection{Interplay with the intensity}

Away from the bistability turning point (TP), the solutions of the Bogoliubov equation are modified as compared to the equilibrium case. 

On the bistability higher branch, the excess of interaction energy over the pump energy $gn > \delta$ results in a splitting of the normal and ghost branches, due to an opening gap whose size increases as the intensity rises. In addition, the dispersion curve recovers a parabolic shape: the regime of collective excitations is lost.

On the bistability lower branch, the positive and negative branches cross each other as the interaction energy drops to $gn < \delta /3 $. This causes the energy of Bogoliubov excitations to become purely imaginary at wavenumbers for which the argument of the square root in Eq.~\eqref{eq:Bog_disp1} becomes negative: 
\begin{equation}
  \delta-3gn < \dfrac{\hbar k^2}{2m_{LP}} < \delta-gn.
  \label{eq:zero_energy}
\end{equation}

If the imaginary part $\mathfrak{Im}(\omega_B(k))$ turns out to be positive, the solutions are unstable. However, the losses $\gamma$ can stabilize the solutions by keeping their imaginary part negative over a large intensity range along the bistability lower branch. In this case, the collective excitations are related to  precursors of instabilities \cite{ciuti_quantum_2005} and can be studied in a steady state regime.

In the following we present an experimental investigation of the dispersion curves and the properties of polariton fluids for these specific situations.

\subsection{Effect of the reservoir on sound propagation}

In order to properly model the behavior of the collective excitations, a dark exciton reservoir has to be considered, as explained in Refs.~\cite{stepanov_dispersion_2019,amelio_reservoir_2020}. This provides an additional two-body interaction channel of energy $\hbar g_{r}n_{r}$, fed by the decay of a fraction of polaritons into non-radiative excitons, at energies well below that of the radiative excitons contributing to the strong coupling. Therefore, as demonstrated in \cite{stepanov_dispersion_2019}, its density $n_r$ scales proportionally to the polariton density $n$. This is the model we will use here, and we will check this behaviour in our experiments.

The reservoir contribution is included in the generalized GPE (\ref{GPE1}) by adding its interaction energy with polaritons. It leads to the following Bogoliubov dispersion relation:
\begin{eqnarray}
  \omega_B(k) &=& \nonumber\\
   &\pm& \sqrt{\left(\frac{\hbar k^2}{2m_{LP}} - \delta + g_{r}n_{r} + 2 gn \right)^2 - (gn)^2} \nonumber\\
 &-& i \dfrac{\gamma}{2}.
  \label{eq:Bog_disp_res}
\end{eqnarray}

Note that, in this expression, the reservoir energy is only added to the average energy of polaritons, fixed by the laser detuning $\delta$, and not to the intrinsic dynamics of the fluid, which is governed by the interaction energy $gn$ between  polaritons. Therefore the reservoir is assumed not to contribute to the dynamics of collective excitations. This simplified picture holds for excitations far from the pump frequency, as disscussed in Ref.~\cite{amelio_reservoir_2020}.
In particular, at the bistability TP, the polariton density is now associated to both the laser and reservoir energy via $\delta = gn + g_{r}n_{r}$. Correspondingly, one calculates the following Bogoliubov equation, same as before:
\begin{equation}
\omega_B(k) = \pm\sqrt{\left(\frac{\hbar k^2}{2m_{LP}}\right)\left(\frac{\hbar k^2}{2m_{LP}} + 2 gn\right)} - i\gamma/2.
\label{Bog_disp2_res}
\end{equation}
At low-k one recovers two linear branches, corresponding to the onset of phonon-like excitations, of speed of sound $c_s^r=\sqrt{\hbar gn/m_{LP}}$ which does not depend on the reservoir energy.

The only significant effect of the reservoir is to lower the renormalization of the system energy due to the interactions between polaritons, with respect to the laser detuning, i.e. $ g n = \delta - g_r n_r$ at the TP. Accordingly, the speed of sound expressed as a function of $\delta$ is also changed by a factor of proportionality $\alpha$ with respect to the speed of sound of the pure polariton fluid:
\begin{equation}
c_s^r=\alpha \hspace{.05cm} c_s = \alpha \sqrt{\frac{\hbar \delta }{m_{LP}}},
\label{reservoir2}
\end{equation}
corresponding to a blueshift contribution
\begin{equation}
g_{r}n_{r} = (1-\alpha^2) \hspace{0.05cm} \delta.
\label{reservoir3}
\end{equation}
We will experimentally verify this model.

\section{Experimental set-up}

Our semiconductor microcavity is made of two highly reflecting planar GaAs/AlGaAs Bragg mirrors, spaced from each other by a 2$\lambda$ thick GaAs substrate to form a high finesse ($\mathcal{F} = 3000$) optical cavity, which embeds three InGaAs quantum wells (QW) at the antinodes of its field. It is placed in a open-flow helium cryostat at the temperature of 4K in order to be able to support robust QW excitons. The strong coupling results in a Rabi energy equal to $\hbar \Omega_R = 5.07$ meV and provides lower polaritons with lifetime of about 15 ps, dominated by the mirror losses.
 
A small wedge between the two mirrors allows to continuously change the detuning between the photons and excitons, from +8 to -4 meV, and consequently the fraction of photons and excitons contributing to the polariton states. It also leads to an energy gradient of the bottom of the lower polariton branch, of about 0.7 $\mu$eV$\cdot \mu$m$^{-1}$ along its axis.

 \begin{figure}
        \includegraphics[width=.75\linewidth]{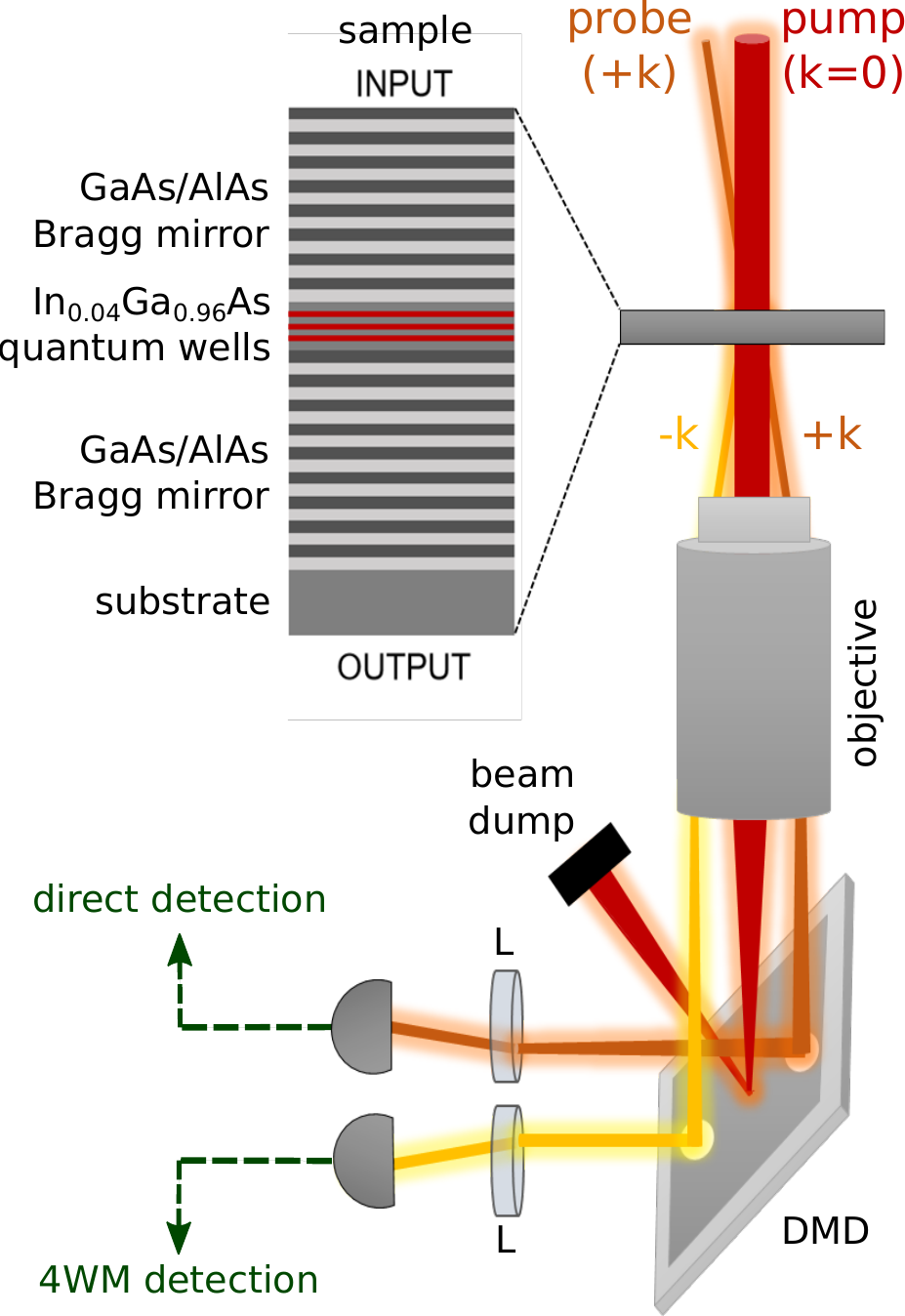}
        \caption{\textbf{Experimental set-up.} The microcavity is placed in a cryostat at 4K. It is illuminated by a pump laser orthogonal to the cavity. The probe laser, modulated at 5 MHz, with an adjustable frequency and in-plane wavevector +k, is detected either in transmission or in reflection (not shown here). When the probe resonates with the pump polariton fluid, it is transmitted and generates an additional 4WM signal at -k. The polariton fluid  is imaged in the reciprocal space on a DMD displaying a pinhole of tunable position, switched between +k and -k. By selecting the +k (-k) angle, a direct (4WM) detection of the probe transmission is operated. The pinhole output is detected by a photodiode, connected to a spectrum analyser demodulating the signal at 5MHz.}
    \label{fig:setup}
\end{figure}

In Fig.~\ref{fig:setup} we show the core of the experimental set-up.
A cw Ti:Sapphire pump laser with a linewidth less than $\SI{1}{\mega\hertz}$ illuminates the cavity at normal incidence in order to create a 2D polariton  fluid at rest in an area of  $100\mu m$, corresponding to the waist of the pump beam. Its frequency is slightly blue detuned from the polariton resonance ($\delta > \sqrt{3}\gamma/2$) in order to operate in the bistable regime, where different densities of the upper and lower branches of the hysteresis loop can be excited by tuning the intensity. The phase and intensity profiles of the pump beam are controlled with a spatial light modulator (SLM).

In order to examine the collective excitations of the fluid, a second  very weak cw probe laser beam is applied on top of the fluid  with a spot of $75\mu m$ in diameter.  Its angle of incidence with respect to the cavity plane is controlled with another SLM, displaying a blazed grating of tunable step size to change the in-plane wavevector $k_{pr}$ of the first diffraction order sent to the sample.

For each selected $k_{pr}$, the probe frequency $\omega_{pr}$ is scanned around the pump frequency $\omega_p$ over a range of 200 GHz ($0.8$ meV), such that only the lower branch of polaritons is excited. When the probe resonates with one of the excitation modes of the fluid, it goes through the sample, resulting in a peak (drop) in the cavity transmission (reflectivity) at the corresponding wavevector and frequency ($k_{pr}$, $\omega_{pr}$), giving direct access to the dispersion curve of collective excitations of the fluid.

To measure such resonances, it is crucial to dissociate the probe signal from the intense background of pump photons transmitted (reflected) by the cavity mirrors. This is achieved by modulating the probe amplitude with an acousto-optic modulator (AOM) before entering the cavity, at the frequency of $f_{\mathrm{mod}}=\SI{5}{\mega\hertz}$, chosen to be significantly higher than the spectral width of the pump laser and its low-frequency noise bandwidth. Then, with a spectrum analyser we perform a demodulation at $f_{\mathrm{mod}}$ of the transmitted (reflected) cavity signals focused on photodiodes, which isolates the probe signal from the pump signal. 
In this way, the spectral resolution of the measurements is only limited by the spectral width of the probe laser, here lower than 1 MHz ($\SI{4}{\nano\eV}$).

Meanwhile, the angular resolution of the detection system is precisely controlled by means of a digital micromirror device (DMD) positioned in the Fourier plane of the cavity. This device displays a tunable radius and position pinhole that acts as a spatial frequency filter before the detection photodiode, providing a high $k$-resolution of $\delta k = \pm\SI{0.0005}{\per\micro\meter}$.

\section{Bogoliubov dispersion curves}

\subsection{Shape of the dispersion curves}

\begin{figure}
        \includegraphics[width=1.0\linewidth]{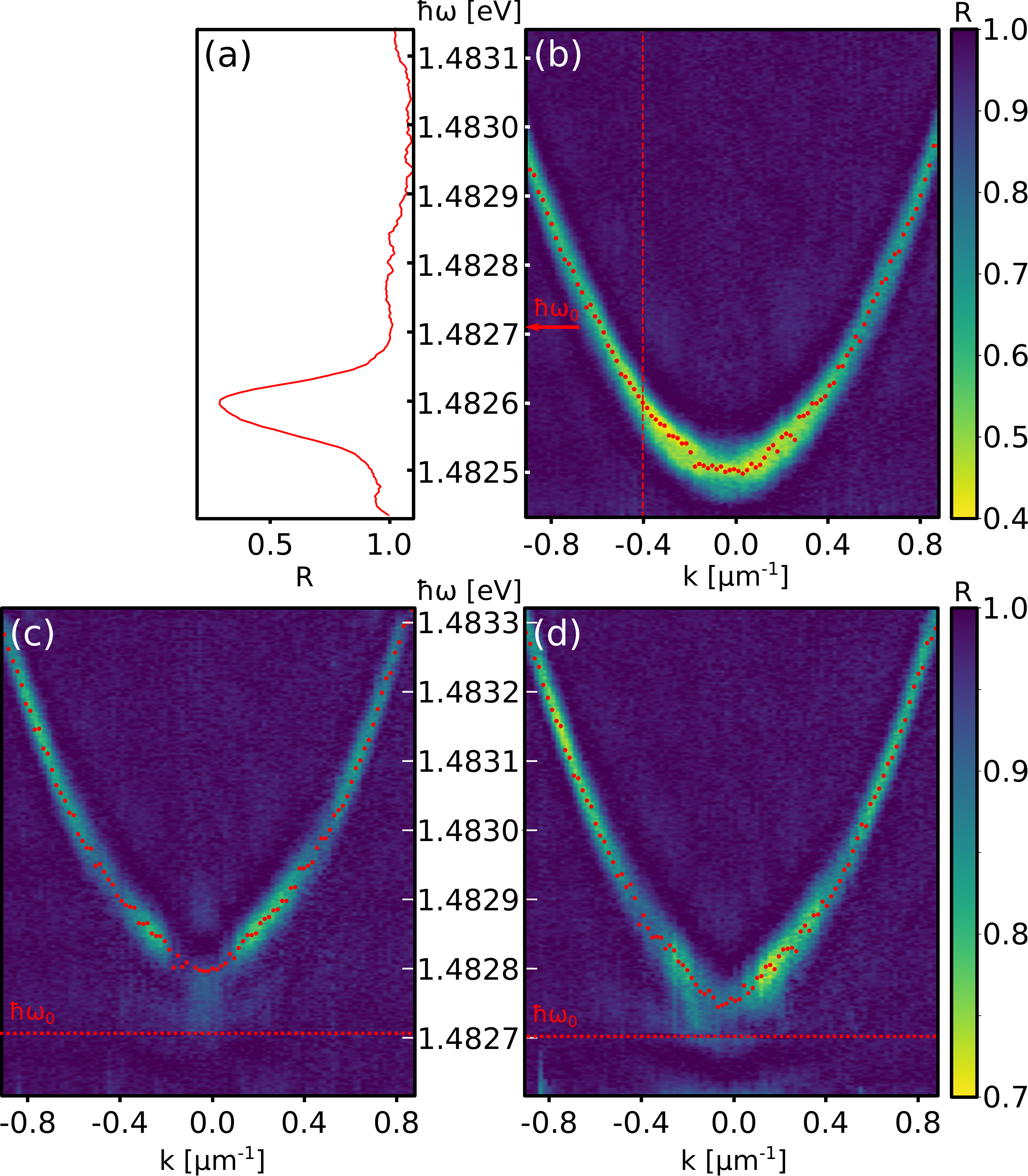}
        \caption{\textbf{Bogoliubov dispersion curves}
        \textbf{(a)} Energy scan of the cavity reflectivity at fixed k = 0.40$\mu$m$^{-1}$ (corresponding to the vertical red dashed line in (b). \textbf{(b)} Dispersion in the linear regime (no pump laser). The red points show the minima of the reflectivity dip for each energy scan. The red arrow indicates the energy $\hbar \omega_0$ of the pump used in panels (b, c). \textbf{(c)} Dispersion in the bistable nonlinear regime (point B in the bistability curve of Fig 1.). \textbf{(d)} Dispersion measured close to the bistability turning point A. The shape of the curve is changed as compared to (c) and the gap is decreased. }
    \label{fig:disp_1}
\end{figure}



 We now study the dispersion relation as a function of the pump intensity $I = |F_p|^2$, for a fixed detuning $\delta=\SI{0.2}{\milli\eV}$ between the pump and the lower polariton energy at $k=0$.


 Fig.~\ref{fig:disp_1} (b-d) shows typical spectra obtained in reflection with the pump probe technique, in the linear and nonlinear regimes. Each successive vertical slice corresponds to a scan of the probe frequency at a given k, separated from each other by $\Delta$k = 0.0189 $\pm$ 0.0005 $\mu$m$^{-1}$. Fig.~\ref{fig:disp_1} (a) gives an example of measured probe signal, with a dip associated to the detection of resonance. For each slice, the energy of the reflectivity minimum gives the real part of the energy of the Bogoliubov dispersion, highlighted by the red dots in Fig. \ref{fig:disp_1} (b-d).
 
 In the linear regime  (Fig.~\ref{fig:disp_1} (b)), the probe scans the cavity without any fluid injected by the pump. We recover the standard dispersion relation of lower polaritons, parabolic over the range of wavenumbers scanned in [-0.95, +0.95] $\mu$m$^{-1}$. From this dispersion we extract the cavity parameters at the considered working point, including the polaritons decay rate given by the full width at half maximun (FWHM) of the reflectivity dips $\hbar\gamma = 80 \mu eV$, and the polariton mass $m_{LP}$ =  5.5$\cdot$10$^{-35}$ kg. The reliability of such a measurement is confirmed by comparison with spectra measured in an out-of-resonance excitation scheme, by detecting the photoluminescence signal of the lower branch with a spectrometer.

 To study the Bogoliubov dispersion of the polariton fluid in non linear regime, we add the pump beam and increase its intensity. As mentioned above, the pump laser frequency is blue detuned from the polariton resonance in order to work in the bistable regime ($\hbar \delta = 0.2$ meV). To operate on the higher branch of the associated hysteresis loop, the laser intensity must first be ramped up until it is above the lower turning point of the bistability (near C in Fig.~\ref{fig:fig1}), where the fluid switches from the low to the high density regime. Then, it is ramped down to address the density region near the TP (points B and A).
 
The dispersion curve shown in Fig. \ref{fig:disp_1} (c), corresponding to point B in Fig. \ref{fig:fig1}  exhibits a blue shift that brings the bottom of the dispersion curve at $k=0$ from 1.4825 to 1.4828 meV, above the pump energy. 

In Fig.~\ref{fig:disp_1} (d), although the laser intensity at its center is close to the bistability turning point, corresponding to point A in Fig. \ref{fig:fig1}, where the linear dispersion should appear at low k, starting from zero detuning as predicted by Eq.~\eqref{eq:Bog_disp1}, the dispersion curve keeps displaying a small gap with respect to the laser frequency and the dispersion is steeper than parabolic, but not fully linear. These effects are investigated in the next section.

\subsection{Spatial dependence of the polariton density}

\begin{figure*}
        \includegraphics[width=.85\textwidth]{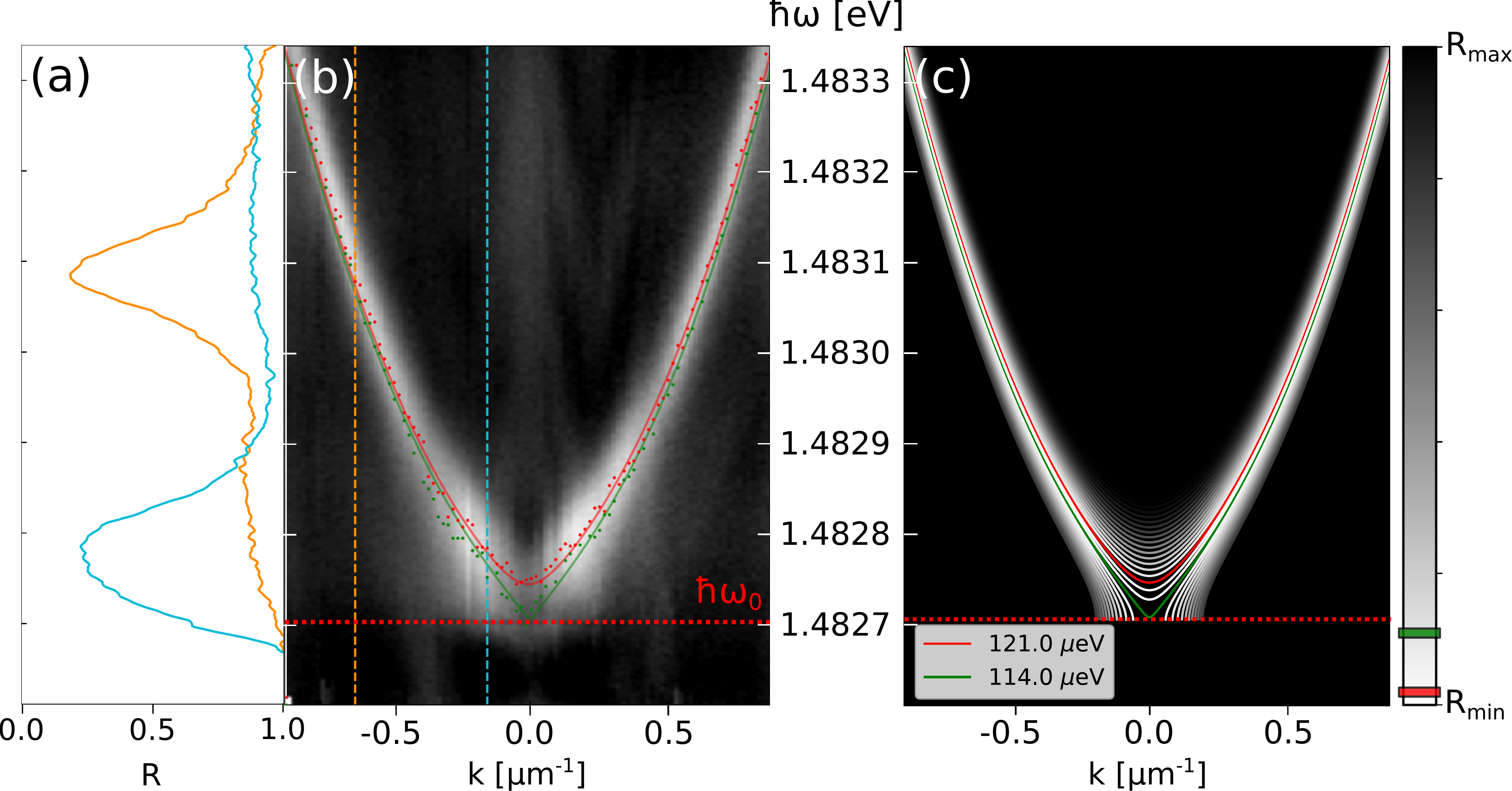}
        \caption{\textbf{Analysis of the shape of the spectrum linewidth.} \textbf{(a)} Vertical slices taken along the blue dashed lines at $k=0.16 \mu m ^{-1}$ and orange dashed lines at $k=0.64 \mu m ^{-1}$ of the spectrum shown in (b). \textbf{(b)} Reflection map of the probe with respect to its wavevector $k$ and energy $\hbar \omega$, at a detuning $\hbar \delta = 0.2$ meV. The horizontal red dotted line indicates the energy of the pump $\hbar \omega_0$. The red points highlight the minima of reflectivity $R_{min}$; the green points the correspond to the reflectivity dips at $R = 0.11 \times \left(R_{max}-R_{min}\right)$. The red and green curves are the corresponding theoretical fits, obtained for $\hbar gn=122\mu eV$ and $\hbar gn=114\mu eV$ respectively. \textbf{(c)} Analytical dispersion curves corresponding to concentric rings of equal width in the Gaussian pump, of reflectivity calculated assuming a Gaussian probe. The red curve corresponds to $\hbar gn=122\mu eV$; the linear green curve to $\hbar gn=114\mu eV$.
}
    \label{fig:disp_3}
\end{figure*}

In Fig.~\ref{fig:disp_1}(d) phonon-like Bogoliubov excitations were not observed because of the Gaussian intensity profile of the laser, which creates a non-uniform polariton density ~\cite{stepanov_dispersion_2019}.
As a result, over the spatial extent excited by the probe beam, collective excitations with different dispersions are scanned and therefore the observed shape of the dispersion curve differs from that expected when considering a uniform pump intensity profile. Specifically, in the present case, gapped dispersions are excited in addition to the gapless one, preventing the observation of a purely sonic dispersion.

In order to solve this problem we made a more detailed analysis of the data, as shown in Fig.~\ref{fig:disp_3}.
In panel \textbf{(b)}, the red dots highlight the minima of the reflectivity dips in the trace of the probe energy scans for each value of wavevector $k$ analysed. They are fitted with Eq.~\eqref{eq:Bog_disp_res} by the red theoretical curve that corresponds to $\hbar gn=122 \mu eV$ (including the reservoir contribution of the reservoir $\hbar g_{r}n_{r}$).
Note that while the FWHM is only 5\% broader than the polariton decay rate at high-$k$, it is much broader at low-$k$, where the Bogoliubov energy has a stronger dependence on $gn$.

To account for this behaviour, we modelled the intensity of the Gaussian pump beam as a set of concentric rings of equal width. Each ring is associated with a discrete $gn$ value, varying from 90 to 156 $\mu eV$, from which we calculate a dispersion curve by using Eq.~\eqref{eq:Bog_disp_res}. These various curves are presented in  Fig.~\ref{fig:disp_3} (c), where the red one corresponds to the red plot fitting the experimental data in Fig.~\ref{fig:disp_3} (b) ($\hbar gn=122 \mu eV$). From this analysis, we see that the dispersion curve changes significantly at low $k$: the fluid switches from gapped parabolic modes at high $g n$ to gapless non-parabolic modes at low $g n$. The linear dispersion highlighted in green is obtained when $\hbar g n$ is equal to 114 $\mu$eV.


Now, in order to reproduce the experimental measurement as closely as possible, it is crucial to weight the contribution of each dispersion of these pump rings to the overall shape of the spectrum, i.e. to assign them a reflectivity when the probe resonates with them. In the experiment, the circular probe spot overlaps with the pump. When the probe has a given frequency $\omega$ and wavevector $k$, it resonates for a specific value of $gn$ given by Eq.~\eqref{eq:Bog_disp_res}. The corresponding ring defined above is thus transmitted ($R=0$). In order to simplify the model, we assume that all the other non-resonant pump rings have a unitary reflectivity ($R=1$). In addition, we take into account the Gaussian distribution of the probe intensity and the area of each ring, as they affect the transmitted and reflected intensity.


Based on these considerations, it is simple to assign an intensity dependent reflectivity $R_{gn}$ to each ring. By comparison with the experimental spectrum in Fig~\ref{fig:disp_3} (b), we can thus in Fig~\ref{fig:disp_3} (c) associate to the red dispersion curve a minimum reflectivity ($R_{122\mu eV} = R_{min}$), and extrapolate the reflectivity of the green sonic dispersion curve, equal to $R_{114 \mu eV} = 0.11 \times \left(R_{max}-R_{min}\right)$, with respect to the maximum $R_{max}$ and minimum $R_{min}$ of the full reflectivity map.



Remarkably, the experimental curve in Fig.~\ref{fig:disp_3}(b) linking the points with a reflectivity of $R = 0.11 \times \left(R_{max}-R_{min}\right)$ reveals a clear linear dispersion at low k, in agreement with the model. This confirms that one can isolate from the experimental spectrum the dispersion corresponding to the bistability TP, calculate its interaction energy and finally extract a speed of sound value.



\subsection{Speed of sound in the fluid}

Thanks to the high resolution in momentum and energy of our technique, we collected precise data on the shape of the polariton spectra depending on the density, and we extracted the dispersion relation corresponding exactly to the turning point of the bistability loop. This was done from the fit of the reflectivity isoline exhibiting at low-k two linear branches, whose slope are equal to the speed of sound. The validity of such a model was confirmed experimentally in Ref.~\cite{claude_2022}, by reshaping the probe into a narrow ring with an SLM in order to measure the spectrum only in the regions of the fluid corresponding to the bistability turning point.

 \begin{figure}[ht]
    \centering
    \includegraphics[width=.75\linewidth]{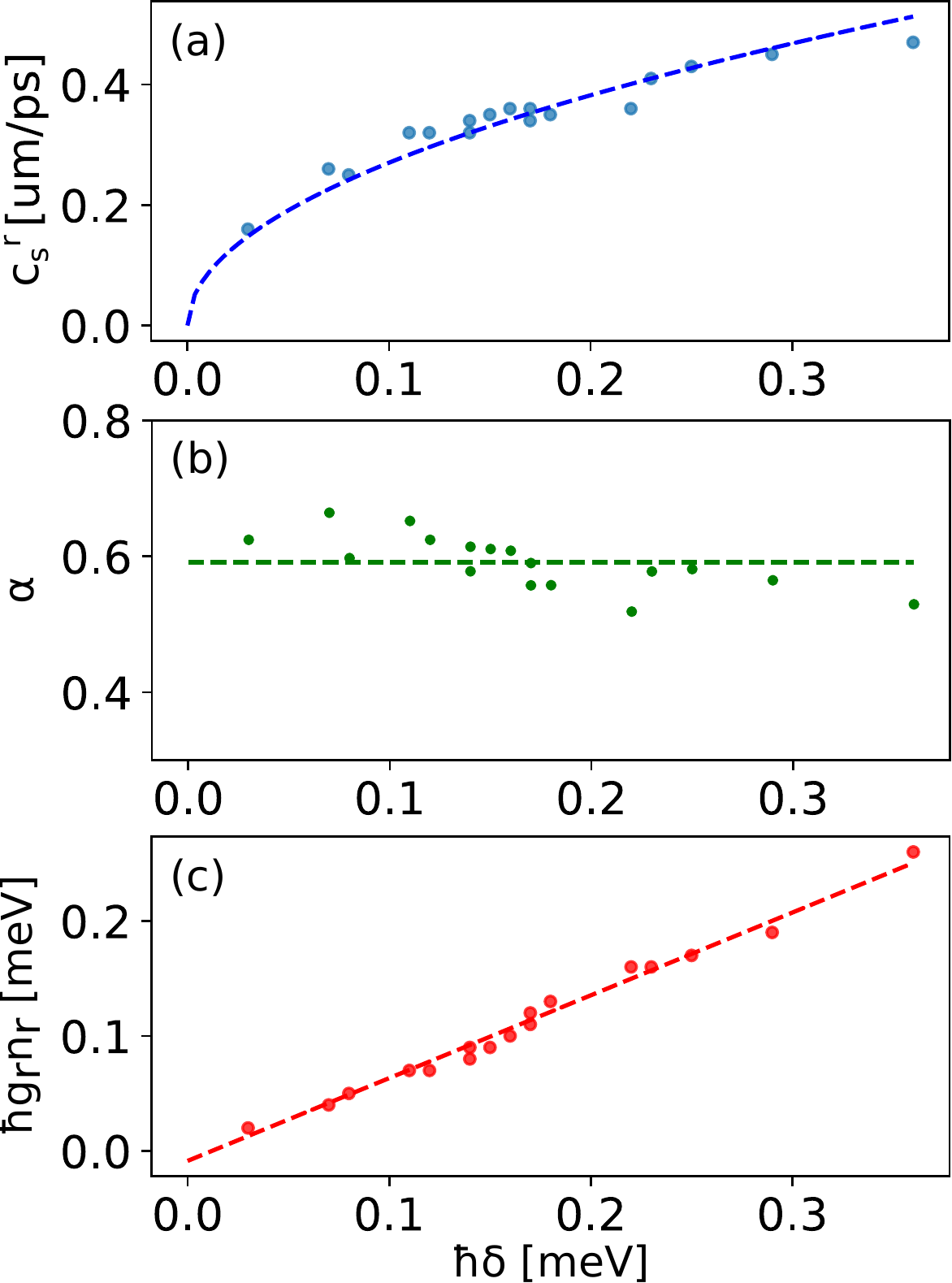}
    
    \caption{Experimental measurements (dots) and fitting (solid line) of \textbf{(a)} the speed of sound $c_s$, \textbf{(b)} the reservoir factor $\alpha$ and \textbf{(c)} the dark exciton reservoir energy $\hbar g_r n_r$, as a function of the pump detuning $\delta$.} 
    \label{fig:cs}
\end{figure}

We measured the speed of sound $c_s$ as a function of the pump detuning $\delta$, by fitting the low-k linear region of the spectra with Eq.~\eqref{Bog_disp2_res}. As shown in Fig.~\ref{fig:cs}(a), we recover the square-root dependence law expected from Eq.~\eqref{reservoir2}. As a result of the effect of the reservoir interactions on the dynamics of the Bogoliubov excitations, the values of the speed of sound are smaller than those predicted by the pure polariton interaction case in Eq.~\eqref{eq:sound}. Rather, they are given by $c_s^r = \alpha \sqrt{\hbar\delta/m_{LP}}$, with $\alpha$ the reservoir factor.


\begin{figure*}
    \centering
    \includegraphics[width=.9\linewidth]{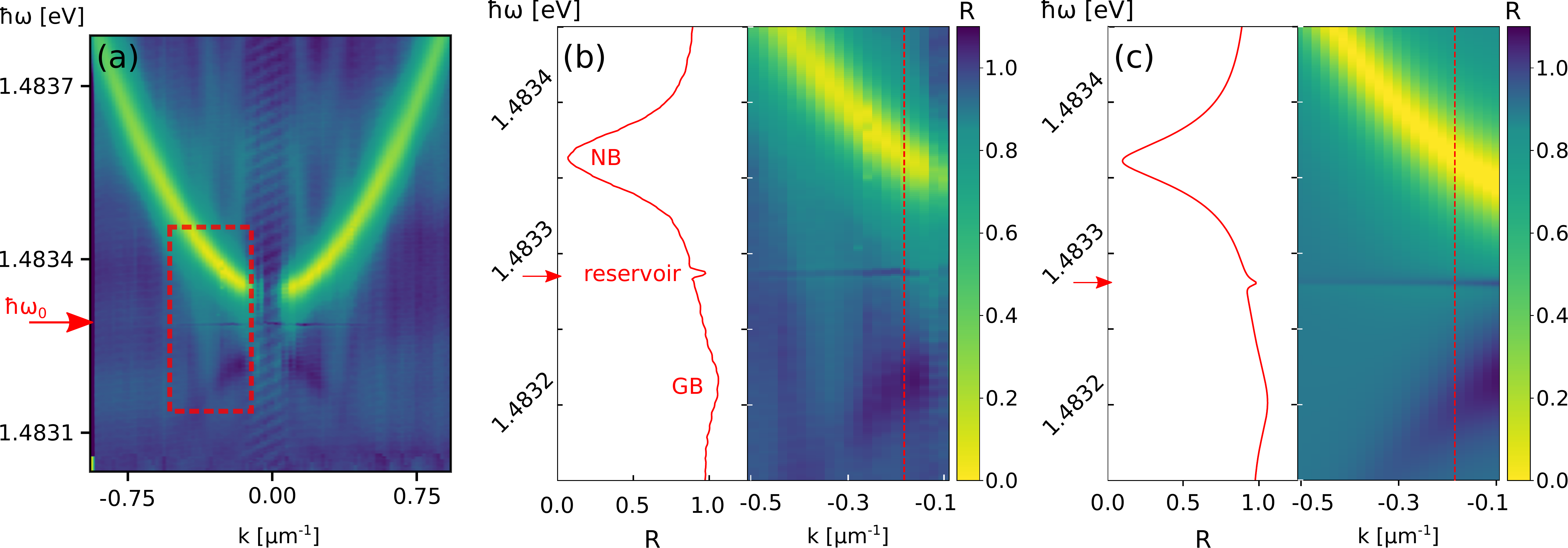}
    \caption{\textbf{Dispersion relation of the exciton reservoir} \textbf{(a)} Full reflection map of the probe. \textbf{(b)} Right panel: zoom in the region of the red frame in (a). Left panel: vertical slice along the red dashed line of the right panel. The normal  (NB) R$<$1 and ghost (GB) R$>$1 branches are observed. The additional narrow flat branch, locally amplified at the pump energy and indicated by the red arrow, corresponds to the dispersion relation of the exciton reservoir, detected at the same energy as the pump. \textbf{(c)} Numerical calculation of the polariton dispersion relation for the same cavity and pump parameters as the experiment.}
    \label{fig:reservoir}
\end{figure*}

From the curve in Fig.~\ref{fig:cs}(a), we measure $\alpha$ plotted in Fig.~\ref{fig:cs}(b). The reservoir factor is nearly constant, in agreement with the theory, with a mean value $\left<\alpha\right> = 0.59$. In addition, Fig.~\ref{fig:cs}(c) shows the energy contribution of the dark exciton reservoir $g_{r}n_{r}$ obtained from the fit of data, which increases linearly with $\delta$ as expected, following a slope of $1-\left<\alpha\right>^2 = 0.65$.

Thus, we obtained precise data on the polariton Bogoliubov dispersion relations, we measured both the speed of sound and the excitonic reservoir contribution, crucial parameters in a wide variety of experiments. Moreover, in the next sections, we show that our technique also provides access to the ghost branches, whose emission is usually several orders of magnitude below the emission of the positive branches, and to a quantitative characterization of the stability of Bogoliubov modes, via the measurement of their linewidth.

\section{Ghost branch and instabilities in the fluid}

 The Bogoliubov dispersion relation includes an additional component: the ghost branch, corresponding to the negative solution $\omega_{B-}$ of Eq.~\eqref{eq:Bog_disp_res}. Unlike the $\omega_{B+}$ branch, its inverted parabolic shape does not match the resonance of the bare optical cavity, making its direct excitation by the pump less efficient, and as a consequence its observation very challenging.
With our set-up, it is possible to increase the detection efficiency by exploiting the four-wave mixing (FWM) process where two pump photons at $\omega_0$ are converted into a pair of photons at $\omega_{B+}$ and $\omega_{B-}$ via the Kerr nonlinearity of polaritons. Indeed, the probe shining the microcavity at $\omega_{B-}$ seeds the scattering of Bogoliubov excitations, from the pump mode toward the ghost branch.

To make this scattering processes as efficient as possible we  have  to increase the density of polaritons excited in the same state by the pump laser, in order to be as close as possible to the regime of single mode FWM. This excitation condition can be reached by using a structured top-hat pump profile, instead of the previous Gaussian one, to obtain a homogeneous fluid density.
This is done by shaping the pump with a second SLM, and comes at the price of a loss in intensity of about 60$\%$ of the incident beam.
Due to the unavoidable intensity, frequency and temperature fluctuations, the top-hat structure makes the system more sensitive to surrounding noise than the Gaussian shape and it results in a rather unstable operation in the vicinity of the sonic point. Therefore, it is not possible to reach exactly the sonic point in that case, preventing the measurement of the speed of sound. However, as we will show this provides an efficient generation of excitations in the ghost branch.

\subsection{Reflectivity experiments}

In Fig.~\ref{fig:reservoir} we present  the typical spectrum obtained in reflection. In panel~(a), for energies higher than that of the pump $\omega > \omega_{0}$, one can see the normal branch as before, with a dip in the probe reflectivity.
In panel (b), for $\omega < \omega_{0}$, the ghost branch manifests itself as an increase in the reflected intensity, above the baseline of the total reflectivity ($R=1$). It is seeded directly by the probe at wavevector k and energy equal to $\omega_{B-}(k)$. It is then amplified by the FWM process that converts photons from the pump toward $\omega_{B-}(k)$ and $\omega_{B+}(-k)$. 
Since the ghost branch is not resonant with the cavity for $k\neq0$, its intensity decreases rapidly with k, whereas it increases when the pump intensity is set as close as possible to the turning point of the bistability, as the phase matching conditions of the FWM become quasi-degenerate, thereby enhancing the efficiency of the scattering processes.

In Fig.~\ref{fig:reservoir} (b), in addition to the aforementioned positive and negative Bogoliubov solutions, a third spectral branch is detected, centered at the pump energy. Its properties, namely its near-zero curvature for the chosen wavevector scale and its very narrow linewidth compared to the standard polariton branches, are typical of those of quantum-well excitons. 
The numerical calculations based on the model proposed in \cite{stepanov_dispersion_2019}, accounting for the relaxation of a fraction of the polariton fluid into a reservoir of dark excitons, and shown in Fig.~\ref{fig:reservoir} (c) accurately reproduce the dispersion relations observed experimentally, and in particular the narrow spectral line at the pump energy, with a full width at half maximum equal to the decay rate set for the excitons in the computation. To match the experimental spectral linewidth, the loss rate of the excitons is fixed at $\hbar \gamma_R$ = 1.5 $\mu$eV, corresponding to a lifetime $\tau$ =  440 ps, consistent with the expected value. 
Therefore, the appearance of such a narrow peak in the dispersion relation in the experiments is a further evidence for the presence of a dark exciton reservoir.

Quite interestingly, the reservoir line corresponds to an increase in the probe reflectivity. On the basis of the model developed in \cite{amelio_reservoir_2020,wouters_goldstone_2007}, this amplification can be understood as the result of stimulated scattering of the pump light on the spatial modulation of the reservoir density due to the interference between the pump and the probe itself.


\subsection{Four wave mixing experiments}

The probe can also be used as an input mode for a four wave mixing (FWM) process generating collective excitations resonant with the normal branch at $\omega_{B+}(k)$, which, by combining with polaritons in the pump mode at $\omega_{0}$, lead to the population of the ghost branch mode at $\omega_{B-}(-k)$. 
The photons emitted from the ghost branch can then be detected in transmission, on the other side of the cavity, by selecting the signal at the wavevector $-k$ opposite to the probe, with the pinhole  of the DMD placed in the reciprocal space. 

\begin{figure*}
    \centering
    \includegraphics[width=1.\linewidth]{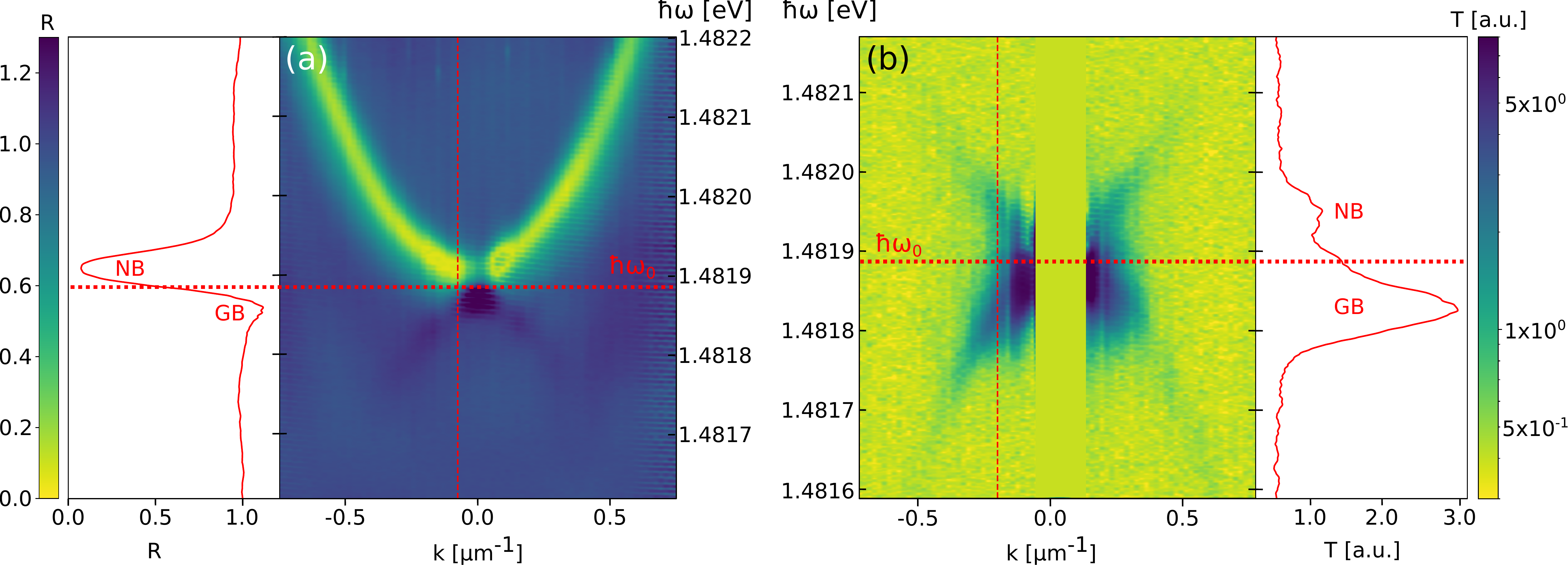}
    \caption{\textbf{Spectra in direct and  FWM detections}, at a pump detuning $\hbar \delta = 0.05$ meV. \textbf{(a)} Right panel: reflection map with the detection set at the wavevector of the probe (direct detection scheme). The pump energy $\hbar \omega_{0}$ is indicated by the horizontal red dotted line. Left panel: slice along the vertical red dashed line of the spectrum. The reflectivity dip R$<$1 corresponds to the normal Bogoliubov branch; the amplification peak R$>$1 to the ghost branch. \textbf{(b)} Left panel: transmission map with the detection set at the opposite wavevector of the probe (indirect detection scheme), under the same experimental conditions as (a). k and $\hbar \omega$ correspond respectively to the wavevenumber and energy emission of the FWM process. Right panel: vertical slice along the vertical red dashed line of the spectrum. Two FWM peaks are detected: the more intense one corresponds to the ghost branch originating from the parametric scattering of the probe at resonance with the normal branch at frequencies above the pump ($\omega >  \omega_{0}$); the other, less intense one to the normal branch, also detected via parametric scattering but originating from the resonance of the probe with the ghost branch at frequencies below the pump ($\omega <  \omega_{0}$). The $\abs{k} <$ 0.15 $\mu$m$^{-1}$ part is masked because the direct transmission angle of the probe is close to the emission angle of the FWM, preventing the filtering of one signal from the other with the DMD.}
    \label{fig:TandR}
\end{figure*}

To illustrate this, Fig.~\ref{fig:TandR}  shows the outcomes of the two different detection schemes, direct and indirect, performed under the same experimental conditions. 
In Fig. \ref{fig:TandR}(a), the detection is operated in reflection, with the DMD selecting the probe wavevector (direct detection): we recover the normal branch in the $R<1$ regions and the seeded ghost branch in the $R>1$ regions. Outside the low-k limit the ghost signal quickly vanishes in the probe reflectivity baseline at $R=1$ .

In Fig. \ref{fig:TandR}(b), the detection is operated in transmission, with the DMD selecting the  wavevector opposite to the probe wavevector (indirect detection). Here, the spectrum is plotted as a function of the wavevector $k$ and energy $\hbar \omega$ of the photons originating from the FWM process: the branch detected below the pump energy corresponds to the ghost branch, resulting from the FWM between the pump polaritons and the probe at resonance with the normal branch; the branch detected above the pump energy corresponds to the normal branch, resulting from the FWM between the pump polaritons and the probe at resonance with the ghost branch.



Interestingly, the signal obtained with this indirect detection scheme is better resolved at large $k$ than that obtained in direct detection.
This method gives excellent access to the properties of the ghost branch. In this case, the measurements were performed for densities in the upper part of the bistability curve. Remarkably, as we show in the next subsection, it also allows to investigate novel properties of the normal and ghost branches in the lower part of the bistability curve.

\subsection{Instabilities in the fluid}

\begin{figure*}[ht!]
    \centering
    \includegraphics[width=1\linewidth]{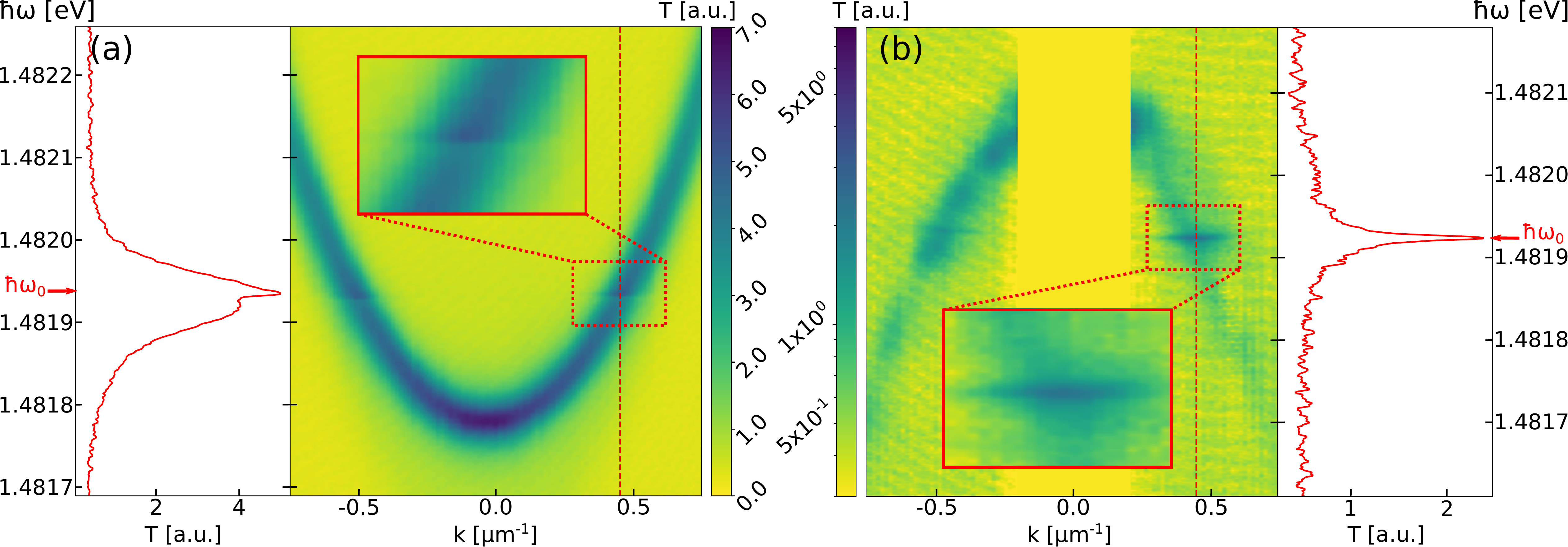}
    \caption{\textbf{Detection of the precursors of dynamical instabilities} for an intensity near the point C of the bistability curve and a detuning $\hbar \delta = 0.16$ meV. \textbf{(a)} Right panel: transmission map of the probe in direct detection. The zoom in the red dashed inset highlights a narrow plateau in the spectrum, signature of the onset of precursor of modulation instabilities at the pump energy $\hbar \omega_0$ indicated by the red arrow. Left panel: slice along the vertical red dashed line of the spectrum. At the pump energy, a narrowing and amplification of the probe is visible, related to the change in the imaginary part of the energy of the collective excitations. \textbf{(b)} Left panel: transmission map of the ghost branch obtained in indirect detection, under the same experimental conditions as (a). $k$ and $\hbar \omega$ are respectively the energy and wavevector of the FWM emission. The ghost branch is detected. The plateau is clearly visible at the crossing point with the normal branch, with a better resolution than in (a). The $\abs{k} <$ 0.25 $\mu$m$^{-1}$ part is masked for the same reason as in Fig. \ref{fig:TandR}. Right panel: slice along the vertical red dashed line of the spectrum. The sub-$\gamma$ linewidth of the resonance peak gives the imaginary part of the energy of the modulation instabilitiy precursor.}
    \label{fig:lowbist_v1}
\end{figure*}

When the excitation of the fluid density is such that $gn < \delta/3$ , condition fulfilled if the laser intensity lies on the lower branch of the bistability, the square root argument in the right part of Eq.~\eqref{eq:Bog_disp1} becomes negative at the wavevectors where the normal and the ghost branches cross each other. Therefore, $\omega_B$ becomes purely imaginary.
If the overall imaginary part is positive, modulational instability modes would appear. However, in the case of our polariton dissipative system, the losses stabilize the fluid by maintaining $\mathfrak{Im}(\omega_B)$ negative over a wide range of driving field intensities.
Fig.~\ref{fig:lowbist_v1}(a) shows the dispersion relation reconstructed under a pump intensity as close as possible to the onset of instability. Remarkably, two plateaus appear at $\omega_0$, exhibiting an increase in the probe transmission and even the onset of some gain as a precursor of the modulational instability.

The observation of the ghost branch in Fig.~\ref{fig:lowbist_v1}(b), is performed in transmission using FWM detection, as in Fig. \ref{fig:TandR}(b). Quite interestingly, note that the FWM in Fig.~\ref{fig:lowbist_v1}(b) does not fulfil the $\omega\leftrightarrow -\omega$ symmetry as only one Bogoliubov branch is clearly visible, namely the one where the incident light drives the normal branch. This unexpected asymmetry of the FWM intensity can be understood by noting that for a sharp-edged top-hat pump in the proximilty of the instability threshold the density of the fluid develops a strong spatial modulation that affects the parametric coupling underlying the ghost branch.

The dispersion curve for the ghost branch is symmetric of the dispersion curve for the normal branch with respect to $\omega_0$, and the instability precursors emerge at the intersection of the positive and negative Bogoliubov solutions. Here, thanks to the FWM detection, only the signal issued from the scattering of polaritons is captured. As a consequence, the cavity resonance, in the background of the fluid, is eliminated from the spectrum, resulting in a better resolution of the very narrow linewidth of the plateaus, without the wide background pedestal of the usual polariton spectral line visible in Fig.~\ref{fig:disp_1}(a).
This narrowing confirms a drastic modification of $\mathfrak{Im}(\omega_B)$ with respect to the standard polariton decay rate $\gamma$, i.e. the precursor of instabilities in the fluid. 

\section{Conclusion}

We have presented an experimental investigation of several new properties of collective excitations in a polariton quantum fluid.
The high energy and wavenumber resolution of our experiment enables the detailed study of the shape of the dispersion curve in the regime of high fluid density and the determination of the speed of sound when operating near the upper turning point of the bistability.
As in \cite{stepanov_dispersion_2019}, we have observed the influence of the intensity profile of the pump on the dispersion at low wavenumber and evidenced the signature of a dark excitonic reservoir that modifies the effective speed of sound in the fluid.
The versatility of our methods is illustrated by the resolution of narrow linewidth plateaus in the dispersion in the regime of low fluid density.

We anticipate that the spectral resolution demonstrated here will enable the investigation of other narrow features such as the excitation of Nambu-Goldstone modes \cite{wouters_goldstone_2007} in this driven-dissipative system.
The study of collective excitations in quantum fluids not only yields insight in their quantum hydrodynamics, but also provides access to the full field theory, which is essential for analogue quantum simulations and the observation of quantum scattering effects such as Hawking radiation or rotational superradiance \cite{jacquet_analogue_2022,braidotti_penrose_2020}, which rely on the engineering of the kinematics of collective excitations of the quantum fluid~\cite{jacquet_influence_2020,jacquet_quantum_2022}.

We acknowledge financial support from the H2020-FETFLAG-2018-2020 project ``PhoQuS'' (n.820392).
MJJ and AB acknowledge support from the \^Ile de France DIM SIRTEQ project ``FOLIAGE'' (n.237992).
IC acknowledges financial support from the Provincia Autonoma di Trento and from the Q@TN initiative.
QG and AB are members of the Institut Universitaire de France.

\bibliography{apssamp.bib}

\providecommand{\noopsort}[1]{}\providecommand{\singleletter}[1]{#1}%
\begin{thebibliography}{30}%
\makeatletter
\providecommand \@ifxundefined [1]{%
 \@ifx{#1\undefined}
}%
\providecommand \@ifnum [1]{%
 \ifnum #1\expandafter \@firstoftwo
 \else \expandafter \@secondoftwo
 \fi
}%
\providecommand \@ifx [1]{%
 \ifx #1\expandafter \@firstoftwo
 \else \expandafter \@secondoftwo
 \fi
}%
\providecommand \natexlab [1]{#1}%
\providecommand \enquote  [1]{``#1''}%
\providecommand \bibnamefont  [1]{#1}%
\providecommand \bibfnamefont [1]{#1}%
\providecommand \citenamefont [1]{#1}%
\providecommand \href@noop [0]{\@secondoftwo}%
\providecommand \href [0]{\begingroup \@sanitize@url \@href}%
\providecommand \@href[1]{\@@startlink{#1}\@@href}%
\providecommand \@@href[1]{\endgroup#1\@@endlink}%
\providecommand \@sanitize@url [0]{\catcode `\\12\catcode `\$12\catcode
  `\&12\catcode `\#12\catcode `\^12\catcode `\_12\catcode `\%12\relax}%
\providecommand \@@startlink[1]{}%
\providecommand \@@endlink[0]{}%
\providecommand \url  [0]{\begingroup\@sanitize@url \@url }%
\providecommand \@url [1]{\endgroup\@href {#1}{\urlprefix }}%
\providecommand \urlprefix  [0]{URL }%
\providecommand \Eprint [0]{\href }%
\providecommand \doibase [0]{https://doi.org/}%
\providecommand \selectlanguage [0]{\@gobble}%
\providecommand \bibinfo  [0]{\@secondoftwo}%
\providecommand \bibfield  [0]{\@secondoftwo}%
\providecommand \translation [1]{[#1]}%
\providecommand \BibitemOpen [0]{}%
\providecommand \bibitemStop [0]{}%
\providecommand \bibitemNoStop [0]{.\EOS\space}%
\providecommand \EOS [0]{\spacefactor3000\relax}%
\providecommand \BibitemShut  [1]{\csname bibitem#1\endcsname}%
\let\auto@bib@innerbib\@empty
\bibitem [{\citenamefont {London}(1938)}]{london_-phenomenon_1938}%
  \BibitemOpen
  \bibfield  {author} {\bibinfo {author} {\bibfnamefont {F.}~\bibnamefont
  {London}},\ }\bibfield  {title} {\bibinfo {title} {The phenomenon of liquid
  helium and the bose-einstein degeneracy},\ }\bibfield  {journal} {\bibinfo
  {journal} {Nature}\ }\textbf {\bibinfo {volume} {141}},\ \href
  {https://doi.org/10.1038/141643a0} {10.1038/141643a0} (\bibinfo {year}
  {1938})\BibitemShut {NoStop}%
\bibitem [{\citenamefont {Landau}(1941)}]{Landau1941}%
  \BibitemOpen
  \bibfield  {author} {\bibinfo {author} {\bibfnamefont {L.}~\bibnamefont
  {Landau}},\ }\bibfield  {title} {\bibinfo {title} {Theory of the
  superfluidity of helium ii},\ }\href {https://doi.org/10.1103/PhysRev.60.356}
  {\bibfield  {journal} {\bibinfo  {journal} {Phys. Rev.}\ }\textbf {\bibinfo
  {volume} {60}},\ \bibinfo {pages} {356} (\bibinfo {year} {1941})}\BibitemShut
  {NoStop}%
\bibitem [{\citenamefont {Bogolyubov}(1947)}]{Bogolyubov:1947zz}%
  \BibitemOpen
  \bibfield  {author} {\bibinfo {author} {\bibfnamefont {N.~N.}\ \bibnamefont
  {Bogolyubov}},\ }\bibfield  {title} {\bibinfo {title} {{On the theory of
  superfluidity}},\ }\href@noop {} {\bibfield  {journal} {\bibinfo  {journal}
  {J. Phys. (USSR)}\ }\textbf {\bibinfo {volume} {11}},\ \bibinfo {pages} {23}
  (\bibinfo {year} {1947})}\BibitemShut {NoStop}%
\bibitem [{\citenamefont {Carusotto}\ and\ \citenamefont
  {Ciuti}(2013)}]{carusotto_quantum_2013}%
  \BibitemOpen
  \bibfield  {author} {\bibinfo {author} {\bibfnamefont {I.}~\bibnamefont
  {Carusotto}}\ and\ \bibinfo {author} {\bibfnamefont {C.}~\bibnamefont
  {Ciuti}},\ }\bibfield  {title} {\bibinfo {title} {Quantum fluids of light},\
  }\bibfield  {journal} {\bibinfo  {journal} {Reviews of Modern Physics}\
  }\textbf {\bibinfo {volume} {85}},\ \href
  {https://doi.org/10.1103/RevModPhys.85.299} {10.1103/RevModPhys.85.299}
  (\bibinfo {year} {2013})\BibitemShut {NoStop}%
\bibitem [{\citenamefont {Jin}\ \emph {et~al.}(1996)\citenamefont {Jin},
  \citenamefont {Ensher}, \citenamefont {Matthews}, \citenamefont {Wieman},\
  and\ \citenamefont {Cornell}}]{jin_collective_1996}%
  \BibitemOpen
  \bibfield  {author} {\bibinfo {author} {\bibfnamefont {D.~S.}\ \bibnamefont
  {Jin}}, \bibinfo {author} {\bibfnamefont {J.~R.}\ \bibnamefont {Ensher}},
  \bibinfo {author} {\bibfnamefont {M.~R.}\ \bibnamefont {Matthews}}, \bibinfo
  {author} {\bibfnamefont {C.~E.}\ \bibnamefont {Wieman}},\ and\ \bibinfo
  {author} {\bibfnamefont {E.~A.}\ \bibnamefont {Cornell}},\ }\bibfield
  {title} {\bibinfo {title} {Collective {Excitations} of a {Bose}-{Einstein}
  {Condensate} in a {Dilute} {Gas}},\ }\href@noop {} {\bibfield  {journal}
  {\bibinfo  {journal} {Physical Review Letters}\ }\textbf {\bibinfo {volume}
  {77}},\ \bibinfo {pages} {4} (\bibinfo {year} {1996})}\BibitemShut {NoStop}%
\bibitem [{\citenamefont {Mewes}\ \emph {et~al.}(1996)\citenamefont {Mewes},
  \citenamefont {Andrews}, \citenamefont {van Druten}, \citenamefont {Kurn},
  \citenamefont {Durfee}, \citenamefont {Townsend},\ and\ \citenamefont
  {Ketterle}}]{mewes_collective_1996}%
  \BibitemOpen
  \bibfield  {author} {\bibinfo {author} {\bibfnamefont {M.-O.}\ \bibnamefont
  {Mewes}}, \bibinfo {author} {\bibfnamefont {M.~R.}\ \bibnamefont {Andrews}},
  \bibinfo {author} {\bibfnamefont {N.~J.}\ \bibnamefont {van Druten}},
  \bibinfo {author} {\bibfnamefont {D.~M.}\ \bibnamefont {Kurn}}, \bibinfo
  {author} {\bibfnamefont {D.~S.}\ \bibnamefont {Durfee}}, \bibinfo {author}
  {\bibfnamefont {C.~G.}\ \bibnamefont {Townsend}},\ and\ \bibinfo {author}
  {\bibfnamefont {W.}~\bibnamefont {Ketterle}},\ }\bibfield  {title} {\bibinfo
  {title} {Collective {Excitations} of a {Bose}-{Einstein} {Condensate} in a
  {Magnetic} {Trap}},\ }\href {https://doi.org/10.1103/PhysRevLett.77.988}
  {\bibfield  {journal} {\bibinfo  {journal} {Physical Review Letters}\
  }\textbf {\bibinfo {volume} {77}},\ \bibinfo {pages} {988} (\bibinfo {year}
  {1996})}\BibitemShut {NoStop}%
\bibitem [{\citenamefont {Onofrio}\ \emph {et~al.}(2000)\citenamefont
  {Onofrio}, \citenamefont {Raman}, \citenamefont {Vogels}, \citenamefont
  {Abo-Shaeer}, \citenamefont {Chikkatur},\ and\ \citenamefont
  {Ketterle}}]{onofrio_observation_2000}%
  \BibitemOpen
  \bibfield  {author} {\bibinfo {author} {\bibfnamefont {R.}~\bibnamefont
  {Onofrio}}, \bibinfo {author} {\bibfnamefont {C.}~\bibnamefont {Raman}},
  \bibinfo {author} {\bibfnamefont {J.~M.}\ \bibnamefont {Vogels}}, \bibinfo
  {author} {\bibfnamefont {J.~R.}\ \bibnamefont {Abo-Shaeer}}, \bibinfo
  {author} {\bibfnamefont {A.~P.}\ \bibnamefont {Chikkatur}},\ and\ \bibinfo
  {author} {\bibfnamefont {W.}~\bibnamefont {Ketterle}},\ }\bibfield  {title}
  {\bibinfo {title} {Observation of {Superfluid} {Flow} in a {Bose}-{Einstein}
  {Condensed} {Gas}},\ }\href {https://doi.org/10.1103/PhysRevLett.85.2228}
  {\bibfield  {journal} {\bibinfo  {journal} {Physical Review Letters}\
  }\textbf {\bibinfo {volume} {85}},\ \bibinfo {pages} {2228} (\bibinfo {year}
  {2000})},\ \bibinfo {note} {publisher: American Physical Society}\BibitemShut
  {NoStop}%
\bibitem [{\citenamefont {Steinhauer}\ \emph {et~al.}(2002)\citenamefont
  {Steinhauer}, \citenamefont {Ozeri}, \citenamefont {Katz},\ and\
  \citenamefont {Davidson}}]{steinhauer_excitation_2002}%
  \BibitemOpen
  \bibfield  {author} {\bibinfo {author} {\bibfnamefont {J.}~\bibnamefont
  {Steinhauer}}, \bibinfo {author} {\bibfnamefont {R.}~\bibnamefont {Ozeri}},
  \bibinfo {author} {\bibfnamefont {N.}~\bibnamefont {Katz}},\ and\ \bibinfo
  {author} {\bibfnamefont {N.}~\bibnamefont {Davidson}},\ }\bibfield  {title}
  {\bibinfo {title} {Excitation {Spectrum} of a {Bose}-{Einstein}
  {Condensate}},\ }\href {https://doi.org/10.1103/PhysRevLett.88.120407}
  {\bibfield  {journal} {\bibinfo  {journal} {Physical Review Letters}\
  }\textbf {\bibinfo {volume} {88}},\ \bibinfo {pages} {120407} (\bibinfo
  {year} {2002})}\BibitemShut {NoStop}%
\bibitem [{\citenamefont {Kasprzak}\ \emph {et~al.}(2006)\citenamefont
  {Kasprzak}, \citenamefont {Richard}, \citenamefont {Kundermann},
  \citenamefont {Baas}, \citenamefont {Jeambrun}, \citenamefont {Keeling},
  \citenamefont {Marchetti}, \citenamefont {Szymańska}, \citenamefont
  {André}, \citenamefont {Staehli}, \citenamefont {Savona}, \citenamefont
  {Littlewood}, \citenamefont {Deveaud},\ and\ \citenamefont
  {Dang}}]{kasprzak_boseeinstein_2006}%
  \BibitemOpen
  \bibfield  {author} {\bibinfo {author} {\bibfnamefont {J.}~\bibnamefont
  {Kasprzak}}, \bibinfo {author} {\bibfnamefont {M.}~\bibnamefont {Richard}},
  \bibinfo {author} {\bibfnamefont {S.}~\bibnamefont {Kundermann}}, \bibinfo
  {author} {\bibfnamefont {A.}~\bibnamefont {Baas}}, \bibinfo {author}
  {\bibfnamefont {P.}~\bibnamefont {Jeambrun}}, \bibinfo {author}
  {\bibfnamefont {J.~M.~J.}\ \bibnamefont {Keeling}}, \bibinfo {author}
  {\bibfnamefont {F.~M.}\ \bibnamefont {Marchetti}}, \bibinfo {author}
  {\bibfnamefont {M.~H.}\ \bibnamefont {Szymańska}}, \bibinfo {author}
  {\bibfnamefont {R.}~\bibnamefont {André}}, \bibinfo {author} {\bibfnamefont
  {J.~L.}\ \bibnamefont {Staehli}}, \bibinfo {author} {\bibfnamefont
  {V.}~\bibnamefont {Savona}}, \bibinfo {author} {\bibfnamefont {P.~B.}\
  \bibnamefont {Littlewood}}, \bibinfo {author} {\bibfnamefont
  {B.}~\bibnamefont {Deveaud}},\ and\ \bibinfo {author} {\bibfnamefont {L.~S.}\
  \bibnamefont {Dang}},\ }\bibfield  {title} {\bibinfo {title}
  {Bose–{Einstein} condensation of exciton polaritons},\ }\bibfield
  {journal} {\bibinfo  {journal} {Nature}\ }\textbf {\bibinfo {volume} {443}},\
  \href {https://doi.org/10.1038/nature05131} {10.1038/nature05131} (\bibinfo
  {year} {2006})\BibitemShut {NoStop}%
\bibitem [{\citenamefont {Balili}\ \emph {et~al.}(2007)\citenamefont {Balili},
  \citenamefont {Hartwell}, \citenamefont {Snoke}, \citenamefont {Pfeiffer},\
  and\ \citenamefont {West}}]{balili_bose-einstein_2007}%
  \BibitemOpen
  \bibfield  {author} {\bibinfo {author} {\bibfnamefont {R.}~\bibnamefont
  {Balili}}, \bibinfo {author} {\bibfnamefont {V.}~\bibnamefont {Hartwell}},
  \bibinfo {author} {\bibfnamefont {D.}~\bibnamefont {Snoke}}, \bibinfo
  {author} {\bibfnamefont {L.}~\bibnamefont {Pfeiffer}},\ and\ \bibinfo
  {author} {\bibfnamefont {K.}~\bibnamefont {West}},\ }\bibfield  {title}
  {\bibinfo {title} {Bose-{Einstein} {Condensation} of {Microcavity}
  {Polaritons} in a {Trap}},\ }\bibfield  {journal} {\bibinfo  {journal}
  {Science}\ }\textbf {\bibinfo {volume} {316}},\ \href
  {https://doi.org/10.1126/science.1140990} {10.1126/science.1140990} (\bibinfo
  {year} {2007})\BibitemShut {NoStop}%
\bibitem [{\citenamefont {Lagoudakis}\ \emph {et~al.}(2008)\citenamefont
  {Lagoudakis}, \citenamefont {Wouters}, \citenamefont {Richard}, \citenamefont
  {Baas}, \citenamefont {Carusotto}, \citenamefont {André}, \citenamefont
  {Dang},\ and\ \citenamefont {Deveaud-Plédran}}]{lagoudakis_quantized_2008}%
  \BibitemOpen
  \bibfield  {author} {\bibinfo {author} {\bibfnamefont {K.~G.}\ \bibnamefont
  {Lagoudakis}}, \bibinfo {author} {\bibfnamefont {M.}~\bibnamefont {Wouters}},
  \bibinfo {author} {\bibfnamefont {M.}~\bibnamefont {Richard}}, \bibinfo
  {author} {\bibfnamefont {A.}~\bibnamefont {Baas}}, \bibinfo {author}
  {\bibfnamefont {I.}~\bibnamefont {Carusotto}}, \bibinfo {author}
  {\bibfnamefont {R.}~\bibnamefont {André}}, \bibinfo {author} {\bibfnamefont
  {L.~S.}\ \bibnamefont {Dang}},\ and\ \bibinfo {author} {\bibfnamefont
  {B.}~\bibnamefont {Deveaud-Plédran}},\ }\bibfield  {title} {\bibinfo {title}
  {Quantized vortices in an exciton–polariton condensate},\ }\bibfield
  {journal} {\bibinfo  {journal} {Nature Physics}\ }\textbf {\bibinfo {volume}
  {4}},\ \href {https://doi.org/10.1038/nphys1051} {10.1038/nphys1051}
  (\bibinfo {year} {2008})\BibitemShut {NoStop}%
\bibitem [{\citenamefont {Utsunomiya}\ \emph {et~al.}(2008)\citenamefont
  {Utsunomiya}, \citenamefont {Tian}, \citenamefont {Roumpos}, \citenamefont
  {Lai}, \citenamefont {Kumada}, \citenamefont {Fujisawa}, \citenamefont
  {Kuwata-Gonokami}, \citenamefont {Löffler}, \citenamefont {Höfling},
  \citenamefont {Forchel},\ and\ \citenamefont
  {Yamamoto}}]{utsunomiya_observation_2008}%
  \BibitemOpen
  \bibfield  {author} {\bibinfo {author} {\bibfnamefont {S.}~\bibnamefont
  {Utsunomiya}}, \bibinfo {author} {\bibfnamefont {L.}~\bibnamefont {Tian}},
  \bibinfo {author} {\bibfnamefont {G.}~\bibnamefont {Roumpos}}, \bibinfo
  {author} {\bibfnamefont {C.~W.}\ \bibnamefont {Lai}}, \bibinfo {author}
  {\bibfnamefont {N.}~\bibnamefont {Kumada}}, \bibinfo {author} {\bibfnamefont
  {T.}~\bibnamefont {Fujisawa}}, \bibinfo {author} {\bibfnamefont
  {M.}~\bibnamefont {Kuwata-Gonokami}}, \bibinfo {author} {\bibfnamefont
  {A.}~\bibnamefont {Löffler}}, \bibinfo {author} {\bibfnamefont
  {S.}~\bibnamefont {Höfling}}, \bibinfo {author} {\bibfnamefont
  {A.}~\bibnamefont {Forchel}},\ and\ \bibinfo {author} {\bibfnamefont
  {Y.}~\bibnamefont {Yamamoto}},\ }\bibfield  {title} {\bibinfo {title}
  {Observation of {Bogoliubov} excitations in exciton-polariton condensates},\
  }\href {https://doi.org/10.1038/nphys1034} {\bibfield  {journal} {\bibinfo
  {journal} {Nature Physics}\ }\textbf {\bibinfo {volume} {4}},\ \bibinfo
  {pages} {700} (\bibinfo {year} {2008})}\BibitemShut {NoStop}%
\bibitem [{\citenamefont {Amo}\ \emph {et~al.}(2009)\citenamefont {Amo},
  \citenamefont {Lefrère}, \citenamefont {Pigeon}, \citenamefont {Adrados},
  \citenamefont {Ciuti}, \citenamefont {Carusotto}, \citenamefont {Houdré},
  \citenamefont {Giacobino},\ and\ \citenamefont
  {Bramati}}]{amo_superfluidity_2009}%
  \BibitemOpen
  \bibfield  {author} {\bibinfo {author} {\bibfnamefont {A.}~\bibnamefont
  {Amo}}, \bibinfo {author} {\bibfnamefont {J.}~\bibnamefont {Lefrère}},
  \bibinfo {author} {\bibfnamefont {S.}~\bibnamefont {Pigeon}}, \bibinfo
  {author} {\bibfnamefont {C.}~\bibnamefont {Adrados}}, \bibinfo {author}
  {\bibfnamefont {C.}~\bibnamefont {Ciuti}}, \bibinfo {author} {\bibfnamefont
  {I.}~\bibnamefont {Carusotto}}, \bibinfo {author} {\bibfnamefont
  {R.}~\bibnamefont {Houdré}}, \bibinfo {author} {\bibfnamefont
  {E.}~\bibnamefont {Giacobino}},\ and\ \bibinfo {author} {\bibfnamefont
  {A.}~\bibnamefont {Bramati}},\ }\bibfield  {title} {\bibinfo {title}
  {Superfluidity of polaritons in semiconductor microcavities},\ }\href
  {https://doi.org/10.1038/nphys1364} {\bibfield  {journal} {\bibinfo
  {journal} {Nature Physics}\ }\textbf {\bibinfo {volume} {5}},\ \bibinfo
  {pages} {805} (\bibinfo {year} {2009})}\BibitemShut {NoStop}%
\bibitem [{\citenamefont {Kohnle}\ \emph {et~al.}(2011)\citenamefont {Kohnle},
  \citenamefont {Léger}, \citenamefont {Wouters}, \citenamefont {Richard},
  \citenamefont {Portella-Oberli},\ and\ \citenamefont
  {Deveaud-Plédran}}]{kohnle_single_2011}%
  \BibitemOpen
  \bibfield  {author} {\bibinfo {author} {\bibfnamefont {V.}~\bibnamefont
  {Kohnle}}, \bibinfo {author} {\bibfnamefont {Y.}~\bibnamefont {Léger}},
  \bibinfo {author} {\bibfnamefont {M.}~\bibnamefont {Wouters}}, \bibinfo
  {author} {\bibfnamefont {M.}~\bibnamefont {Richard}}, \bibinfo {author}
  {\bibfnamefont {M.~T.}\ \bibnamefont {Portella-Oberli}},\ and\ \bibinfo
  {author} {\bibfnamefont {B.}~\bibnamefont {Deveaud-Plédran}},\ }\bibfield
  {title} {\bibinfo {title} {From {Single} {Particle} to {Superfluid}
  {Excitations} in a {Dissipative} {Polariton} {Gas}},\ }\href
  {https://doi.org/10.1103/PhysRevLett.106.255302} {\bibfield  {journal}
  {\bibinfo  {journal} {Physical Review Letters}\ }\textbf {\bibinfo {volume}
  {106}},\ \bibinfo {pages} {255302} (\bibinfo {year} {2011})}\BibitemShut
  {NoStop}%
\bibitem [{\citenamefont {Stepanov}\ \emph {et~al.}(2019)\citenamefont
  {Stepanov}, \citenamefont {Amelio}, \citenamefont {Rousset}, \citenamefont
  {Bloch}, \citenamefont {Lemaître}, \citenamefont {Amo}, \citenamefont
  {Minguzzi}, \citenamefont {Carusotto},\ and\ \citenamefont
  {Richard}}]{stepanov_dispersion_2019}%
  \BibitemOpen
  \bibfield  {author} {\bibinfo {author} {\bibfnamefont {P.}~\bibnamefont
  {Stepanov}}, \bibinfo {author} {\bibfnamefont {I.}~\bibnamefont {Amelio}},
  \bibinfo {author} {\bibfnamefont {J.-G.}\ \bibnamefont {Rousset}}, \bibinfo
  {author} {\bibfnamefont {J.}~\bibnamefont {Bloch}}, \bibinfo {author}
  {\bibfnamefont {A.}~\bibnamefont {Lemaître}}, \bibinfo {author}
  {\bibfnamefont {A.}~\bibnamefont {Amo}}, \bibinfo {author} {\bibfnamefont
  {A.}~\bibnamefont {Minguzzi}}, \bibinfo {author} {\bibfnamefont
  {I.}~\bibnamefont {Carusotto}},\ and\ \bibinfo {author} {\bibfnamefont
  {M.}~\bibnamefont {Richard}},\ }\bibfield  {title} {\bibinfo {title}
  {Dispersion relation of the collective excitations in a resonantly driven
  polariton fluid},\ }\href {https://doi.org/10.1038/s41467-019-11886-3}
  {\bibfield  {journal} {\bibinfo  {journal} {Nature Communications}\ }\textbf
  {\bibinfo {volume} {10}},\ \bibinfo {pages} {3869} (\bibinfo {year}
  {2019})}\BibitemShut {NoStop}%
\bibitem [{\citenamefont {Pieczarka}\ \emph {et~al.}(2020)\citenamefont
  {Pieczarka}, \citenamefont {Estrecho}, \citenamefont {Boozarjmehr},
  \citenamefont {Bleu}, \citenamefont {Steger}, \citenamefont {West},
  \citenamefont {Pfeiffer}, \citenamefont {Snoke}, \citenamefont {Levinsen},
  \citenamefont {Parish}, \citenamefont {Truscott},\ and\ \citenamefont
  {Ostrovskaya}}]{pieczarka_observation_2020}%
  \BibitemOpen
  \bibfield  {author} {\bibinfo {author} {\bibfnamefont {M.}~\bibnamefont
  {Pieczarka}}, \bibinfo {author} {\bibfnamefont {E.}~\bibnamefont {Estrecho}},
  \bibinfo {author} {\bibfnamefont {M.}~\bibnamefont {Boozarjmehr}}, \bibinfo
  {author} {\bibfnamefont {O.}~\bibnamefont {Bleu}}, \bibinfo {author}
  {\bibfnamefont {M.}~\bibnamefont {Steger}}, \bibinfo {author} {\bibfnamefont
  {K.}~\bibnamefont {West}}, \bibinfo {author} {\bibfnamefont {L.~N.}\
  \bibnamefont {Pfeiffer}}, \bibinfo {author} {\bibfnamefont {D.~W.}\
  \bibnamefont {Snoke}}, \bibinfo {author} {\bibfnamefont {J.}~\bibnamefont
  {Levinsen}}, \bibinfo {author} {\bibfnamefont {M.~M.}\ \bibnamefont
  {Parish}}, \bibinfo {author} {\bibfnamefont {A.~G.}\ \bibnamefont
  {Truscott}},\ and\ \bibinfo {author} {\bibfnamefont {E.~A.}\ \bibnamefont
  {Ostrovskaya}},\ }\bibfield  {title} {\bibinfo {title} {Observation of
  quantum depletion in a non-equilibrium exciton–polariton condensate},\
  }\href {https://doi.org/10.1038/s41467-019-14243-6} {\bibfield  {journal}
  {\bibinfo  {journal} {Nature Communications}\ }\textbf {\bibinfo {volume}
  {11}},\ \bibinfo {pages} {429} (\bibinfo {year} {2020})}\BibitemShut
  {NoStop}%
\bibitem [{\citenamefont {Claude}\ \emph {et~al.}(2022)\citenamefont {Claude},
  \citenamefont {Jacquet}, \citenamefont {Usciati}, \citenamefont {Carusotto},
  \citenamefont {Giacobino}, \citenamefont {Bramati},\ and\ \citenamefont
  {Glorieux}}]{claude_2022}%
  \BibitemOpen
  \bibfield  {author} {\bibinfo {author} {\bibfnamefont {F.}~\bibnamefont
  {Claude}}, \bibinfo {author} {\bibfnamefont {M.}~\bibnamefont {Jacquet}},
  \bibinfo {author} {\bibfnamefont {R.}~\bibnamefont {Usciati}}, \bibinfo
  {author} {\bibfnamefont {I.}~\bibnamefont {Carusotto}}, \bibinfo {author}
  {\bibfnamefont {E.}~\bibnamefont {Giacobino}}, \bibinfo {author}
  {\bibfnamefont {A.}~\bibnamefont {Bramati}},\ and\ \bibinfo {author}
  {\bibfnamefont {Q.}~\bibnamefont {Glorieux}},\ }\bibfield  {title} {\bibinfo
  {title} {High-{Resolution} {Coherent} {Probe} {Spectroscopy} of a {Polariton}
  {Quantum} {Fluid}},\ }\href {https://doi.org/10.1103/PhysRevLett.129.103601}
  {\bibfield  {journal} {\bibinfo  {journal} {Physical Review Letters}\
  }\textbf {\bibinfo {volume} {129}},\ \bibinfo {pages} {103601} (\bibinfo
  {year} {2022})}\BibitemShut {NoStop}%
\bibitem [{\citenamefont {Amelio}\ and\ \citenamefont
  {Carusotto}(2020)}]{amelio_perspectives_2020}%
  \BibitemOpen
  \bibfield  {author} {\bibinfo {author} {\bibfnamefont {I.}~\bibnamefont
  {Amelio}}\ and\ \bibinfo {author} {\bibfnamefont {I.}~\bibnamefont
  {Carusotto}},\ }\bibfield  {title} {\bibinfo {title} {Perspectives in
  superfluidity in resonantly driven polariton fluids},\ }\href
  {https://doi.org/10.1103/PhysRevB.101.064505} {\bibfield  {journal} {\bibinfo
   {journal} {Physical Review B}\ }\textbf {\bibinfo {volume} {101}},\ \bibinfo
  {pages} {064505} (\bibinfo {year} {2020})}\BibitemShut {NoStop}%
\bibitem [{\citenamefont {Kohnle}\ \emph {et~al.}(2012)\citenamefont {Kohnle},
  \citenamefont {Léger}, \citenamefont {Wouters}, \citenamefont {Richard},
  \citenamefont {Portella-Oberli},\ and\ \citenamefont
  {Deveaud}}]{kohnle_four-wave_2012}%
  \BibitemOpen
  \bibfield  {author} {\bibinfo {author} {\bibfnamefont {V.}~\bibnamefont
  {Kohnle}}, \bibinfo {author} {\bibfnamefont {Y.}~\bibnamefont {Léger}},
  \bibinfo {author} {\bibfnamefont {M.}~\bibnamefont {Wouters}}, \bibinfo
  {author} {\bibfnamefont {M.}~\bibnamefont {Richard}}, \bibinfo {author}
  {\bibfnamefont {M.~T.}\ \bibnamefont {Portella-Oberli}},\ and\ \bibinfo
  {author} {\bibfnamefont {B.}~\bibnamefont {Deveaud}},\ }\bibfield  {title}
  {\bibinfo {title} {Four-wave mixing excitations in a dissipative polariton
  quantum fluid},\ }\href {https://doi.org/10.1103/PhysRevB.86.064508}
  {\bibfield  {journal} {\bibinfo  {journal} {Physical Review B}\ }\textbf
  {\bibinfo {volume} {86}},\ \bibinfo {pages} {064508} (\bibinfo {year}
  {2012})}\BibitemShut {NoStop}%
\bibitem [{\citenamefont {Sermage}\ \emph {et~al.}(1996)\citenamefont
  {Sermage}, \citenamefont {Long}, \citenamefont {Abram}, \citenamefont
  {Marzin}, \citenamefont {Bloch}, \citenamefont {Planel},\ and\ \citenamefont
  {Thierry-Mieg}}]{sermage_lifetime_1996}%
  \BibitemOpen
  \bibfield  {author} {\bibinfo {author} {\bibfnamefont {B.}~\bibnamefont
  {Sermage}}, \bibinfo {author} {\bibfnamefont {S.}~\bibnamefont {Long}},
  \bibinfo {author} {\bibfnamefont {I.}~\bibnamefont {Abram}}, \bibinfo
  {author} {\bibfnamefont {J.~Y.}\ \bibnamefont {Marzin}}, \bibinfo {author}
  {\bibfnamefont {J.}~\bibnamefont {Bloch}}, \bibinfo {author} {\bibfnamefont
  {R.}~\bibnamefont {Planel}},\ and\ \bibinfo {author} {\bibfnamefont
  {V.}~\bibnamefont {Thierry-Mieg}},\ }\bibfield  {title} {\bibinfo {title}
  {Time-resolved spontaneous emission of excitons in a microcavity: Behavior of
  the individual exciton-photon mixed states},\ }\href
  {https://doi.org/10.1103/PhysRevB.53.16516} {\bibfield  {journal} {\bibinfo
  {journal} {Phys. Rev. B}\ }\textbf {\bibinfo {volume} {53}},\ \bibinfo
  {pages} {16516} (\bibinfo {year} {1996})}\BibitemShut {NoStop}%
\bibitem [{\citenamefont {Bloch}\ and\ \citenamefont
  {Marzin}(1997)}]{Bloch_lifetime_1997}%
  \BibitemOpen
  \bibfield  {author} {\bibinfo {author} {\bibfnamefont {J.}~\bibnamefont
  {Bloch}}\ and\ \bibinfo {author} {\bibfnamefont {J.~Y.}\ \bibnamefont
  {Marzin}},\ }\bibfield  {title} {\bibinfo {title} {Photoluminescence dynamics
  of cavity polaritons under resonant excitation in the picosecond range},\
  }\href {https://doi.org/10.1103/PhysRevB.56.2103} {\bibfield  {journal}
  {\bibinfo  {journal} {Phys. Rev. B}\ }\textbf {\bibinfo {volume} {56}},\
  \bibinfo {pages} {2103} (\bibinfo {year} {1997})}\BibitemShut {NoStop}%
\bibitem [{\citenamefont {Ciuti}\ and\ \citenamefont
  {Carusotto}(2005)}]{ciuti_quantum_2005}%
  \BibitemOpen
  \bibfield  {author} {\bibinfo {author} {\bibfnamefont {C.}~\bibnamefont
  {Ciuti}}\ and\ \bibinfo {author} {\bibfnamefont {I.}~\bibnamefont
  {Carusotto}},\ }\bibfield  {title} {\bibinfo {title} {Quantum fluid effects
  and parametric instabilities in microcavities},\ }\href
  {https://doi.org/10.1002/pssb.200560961} {\bibfield  {journal} {\bibinfo
  {journal} {physica status solidi (b)}\ }\textbf {\bibinfo {volume} {242}},\
  \bibinfo {pages} {2224} (\bibinfo {year} {2005})}\BibitemShut {NoStop}%
\bibitem [{\citenamefont {Baas}\ \emph {et~al.}(2004)\citenamefont {Baas},
  \citenamefont {Karr}, \citenamefont {Eleuch},\ and\ \citenamefont
  {Giacobino}}]{baas_bista_2004}%
  \BibitemOpen
  \bibfield  {author} {\bibinfo {author} {\bibfnamefont {A.}~\bibnamefont
  {Baas}}, \bibinfo {author} {\bibfnamefont {J.~P.}\ \bibnamefont {Karr}},
  \bibinfo {author} {\bibfnamefont {H.}~\bibnamefont {Eleuch}},\ and\ \bibinfo
  {author} {\bibfnamefont {E.}~\bibnamefont {Giacobino}},\ }\bibfield  {title}
  {\bibinfo {title} {Optical bistability in semiconductor microcavities},\
  }\href {https://doi.org/10.1103/PhysRevA.69.023809} {\bibfield  {journal}
  {\bibinfo  {journal} {Phys. Rev. A}\ }\textbf {\bibinfo {volume} {69}},\
  \bibinfo {pages} {023809} (\bibinfo {year} {2004})}\BibitemShut {NoStop}%
\bibitem [{\citenamefont {Hakim}\ \emph {et~al.}(2022)\citenamefont {Hakim},
  \citenamefont {Pigeon},\ and\ \citenamefont
  {Aftalion}}]{hakim_metamorphosis_2022}%
  \BibitemOpen
  \bibfield  {author} {\bibinfo {author} {\bibfnamefont {V.}~\bibnamefont
  {Hakim}}, \bibinfo {author} {\bibfnamefont {S.}~\bibnamefont {Pigeon}},\ and\
  \bibinfo {author} {\bibfnamefont {A.}~\bibnamefont {Aftalion}},\ }\bibfield
  {title} {\bibinfo {title} {Metamorphosis of the {Landau} transition in the
  flow of a resonantly-driven bistable polariton fluid},\ }\bibfield  {journal}
  {\bibinfo  {journal} {arXiv:2203.11351 [cond-mat, physics:nlin]}\ }\href
  {https://doi.org/https://doi.org/10.48550/arXiv.2203.11351}
  {https://doi.org/10.48550/arXiv.2203.11351} (\bibinfo {year}
  {2022})\BibitemShut {NoStop}%
\bibitem [{\citenamefont {Amelio}\ \emph {et~al.}(2020)\citenamefont {Amelio},
  \citenamefont {Minguzzi}, \citenamefont {Richard},\ and\ \citenamefont
  {Carusotto}}]{amelio_reservoir_2020}%
  \BibitemOpen
  \bibfield  {author} {\bibinfo {author} {\bibfnamefont {I.}~\bibnamefont
  {Amelio}}, \bibinfo {author} {\bibfnamefont {A.}~\bibnamefont {Minguzzi}},
  \bibinfo {author} {\bibfnamefont {M.}~\bibnamefont {Richard}},\ and\ \bibinfo
  {author} {\bibfnamefont {I.}~\bibnamefont {Carusotto}},\ }\bibfield  {title}
  {\bibinfo {title} {Galilean boosts and superfluidity of resonantly driven
  polariton fluids in the presence of an incoherent reservoir},\ }\href
  {https://doi.org/10.1103/PhysRevResearch.2.023158} {\bibfield  {journal}
  {\bibinfo  {journal} {Phys. Rev. Research}\ }\textbf {\bibinfo {volume}
  {2}},\ \bibinfo {pages} {023158} (\bibinfo {year} {2020})}\BibitemShut
  {NoStop}%
\bibitem [{\citenamefont {Wouters}\ and\ \citenamefont
  {Carusotto}(2007)}]{wouters_goldstone_2007}%
  \BibitemOpen
  \bibfield  {author} {\bibinfo {author} {\bibfnamefont {M.}~\bibnamefont
  {Wouters}}\ and\ \bibinfo {author} {\bibfnamefont {I.}~\bibnamefont
  {Carusotto}},\ }\bibfield  {title} {\bibinfo {title} {Goldstone mode of
  optical parametric oscillators in planar semiconductor microcavities in the
  strong-coupling regime},\ }\href {https://doi.org/10.1103/PhysRevA.76.043807}
  {\bibfield  {journal} {\bibinfo  {journal} {Physical Review A}\ }\textbf
  {\bibinfo {volume} {76}},\ \bibinfo {pages} {043807} (\bibinfo {year}
  {2007})}\BibitemShut {NoStop}%
\bibitem [{\citenamefont {Jacquet}\ \emph
  {et~al.}(2022{\natexlab{a}})\citenamefont {Jacquet}, \citenamefont {Joly},
  \citenamefont {Claude}, \citenamefont {Giacomelli}, \citenamefont {Glorieux},
  \citenamefont {Bramati}, \citenamefont {Carusotto},\ and\ \citenamefont
  {Giacobino}}]{jacquet_analogue_2022}%
  \BibitemOpen
  \bibfield  {author} {\bibinfo {author} {\bibfnamefont {M.~J.}\ \bibnamefont
  {Jacquet}}, \bibinfo {author} {\bibfnamefont {M.}~\bibnamefont {Joly}},
  \bibinfo {author} {\bibfnamefont {F.}~\bibnamefont {Claude}}, \bibinfo
  {author} {\bibfnamefont {L.}~\bibnamefont {Giacomelli}}, \bibinfo {author}
  {\bibfnamefont {Q.}~\bibnamefont {Glorieux}}, \bibinfo {author}
  {\bibfnamefont {A.}~\bibnamefont {Bramati}}, \bibinfo {author} {\bibfnamefont
  {I.}~\bibnamefont {Carusotto}},\ and\ \bibinfo {author} {\bibfnamefont
  {E.}~\bibnamefont {Giacobino}},\ }\bibfield  {title} {\bibinfo {title}
  {Analogue quantum simulation of the {Hawking} effect in a polariton
  superfluid},\ }\href {https://doi.org/10.1140/epjd/s10053-022-00477-5}
  {\bibfield  {journal} {\bibinfo  {journal} {The European Physical Journal D}\
  }\textbf {\bibinfo {volume} {76}},\ \bibinfo {pages} {152} (\bibinfo {year}
  {2022}{\natexlab{a}})}\BibitemShut {NoStop}%
\bibitem [{\citenamefont {Braidotti}\ \emph {et~al.}(2020)\citenamefont
  {Braidotti}, \citenamefont {Faccio},\ and\ \citenamefont
  {Wright}}]{braidotti_penrose_2020}%
  \BibitemOpen
  \bibfield  {author} {\bibinfo {author} {\bibfnamefont {M.~C.}\ \bibnamefont
  {Braidotti}}, \bibinfo {author} {\bibfnamefont {D.}~\bibnamefont {Faccio}},\
  and\ \bibinfo {author} {\bibfnamefont {E.~M.}\ \bibnamefont {Wright}},\
  }\bibfield  {title} {\bibinfo {title} {Penrose {Superradiance} in {Nonlinear}
  {Optics}},\ }\href {https://doi.org/10.1103/PhysRevLett.125.193902}
  {\bibfield  {journal} {\bibinfo  {journal} {Physical Review Letters}\
  }\textbf {\bibinfo {volume} {125}},\ \bibinfo {pages} {193902} (\bibinfo
  {year} {2020})}\BibitemShut {NoStop}%
\bibitem [{\citenamefont {Jacquet}\ and\ \citenamefont
  {K\"onig}(2020)}]{jacquet_influence_2020}%
  \BibitemOpen
  \bibfield  {author} {\bibinfo {author} {\bibfnamefont {M.~J.}\ \bibnamefont
  {Jacquet}}\ and\ \bibinfo {author} {\bibfnamefont {F.}~\bibnamefont
  {K\"onig}},\ }\bibfield  {title} {\bibinfo {title} {The influence of
  spacetime curvature on quantum emission in optical analogues to gravity},\
  }\href {https://doi.org/10.21468/SciPostPhysCore.3.1.005} {\bibfield
  {journal} {\bibinfo  {journal} {SciPost Physics Core}\ }\textbf {\bibinfo
  {volume} {3}},\ \bibinfo {pages} {005} (\bibinfo {year} {2020})}\BibitemShut
  {NoStop}%
\bibitem [{\citenamefont {Jacquet}\ \emph
  {et~al.}(2022{\natexlab{b}})\citenamefont {Jacquet}, \citenamefont
  {Giacomelli}, \citenamefont {Valnais}, \citenamefont {Joly}, \citenamefont
  {Claude}, \citenamefont {Giacobino}, \citenamefont {Glorieux}, \citenamefont
  {Carusotto},\ and\ \citenamefont {Bramati}}]{jacquet_quantum_2022}%
  \BibitemOpen
  \bibfield  {author} {\bibinfo {author} {\bibfnamefont {M.~J.}\ \bibnamefont
  {Jacquet}}, \bibinfo {author} {\bibfnamefont {L.}~\bibnamefont {Giacomelli}},
  \bibinfo {author} {\bibfnamefont {Q.}~\bibnamefont {Valnais}}, \bibinfo
  {author} {\bibfnamefont {M.}~\bibnamefont {Joly}}, \bibinfo {author}
  {\bibfnamefont {F.}~\bibnamefont {Claude}}, \bibinfo {author} {\bibfnamefont
  {E.}~\bibnamefont {Giacobino}}, \bibinfo {author} {\bibfnamefont
  {Q.}~\bibnamefont {Glorieux}}, \bibinfo {author} {\bibfnamefont
  {I.}~\bibnamefont {Carusotto}},\ and\ \bibinfo {author} {\bibfnamefont
  {A.}~\bibnamefont {Bramati}},\ }\bibfield  {title} {\bibinfo {title} {Quantum
  vacuum excitation of a quasi-normal mode in an analog model of black hole
  spacetime},\ }\href@noop {} {\bibfield  {journal} {\bibinfo  {journal}
  {arXiv:2110.14452}\ } (\bibinfo {year} {2022}{\natexlab{b}})}\BibitemShut
  {NoStop}%
\end{thebibliography}%

\end{document}